\documentclass{JFM-FLM_Au}

\lefttitle{Karban}
\righttitle{Journal of Fluid Mechanics}

\title{Pressure-based separation of sources and sinks in the nonlinear forcing of turbulent channel flow}

\author{Ugur Karban\corresp{\email{ukarban@metu.edu.tr}}}
\affiliation{Aerospace Engineering Department, Middle East Technical University,
	Ankara, Turkey}
\corresau{Ugur Karban, ukarban@metu.edu.tr}

\newcommand{\ii}{\mathrm{i}}

\newcommand{\Hu}{\mathsfbi{H}_{\mathrm{q}}}
\newcommand{\Hp}{\mathsfbi{H}_{\mathrm{p}}}
\newcommand{\Gp}{\mathsfbi{G}_{\mathrm{p}}}
\newcommand{\Dmat}{\mathsfbi{D}}
\newcommand{\Bs}{\mathsfbi{B}_{\mathrm{s}}}
\newcommand{\Wmat}{\mathsfbi{W}}
\newcommand{\Js}{\mathsfbi{J}_{\mathrm{s}}}
\newcommand{\Rs}{\mathsfbi{R}_{\mathrm{s}}}
\newcommand{\Uj}{\mathsfbi{U}_{J}}
\newcommand{\Sa}{\mathsfbi{S}_{\mathrm{a}}}
\newcommand{\Na}{\mathsfbi{N}_{\mathrm{a}}}
\newcommand{\Nz}{\mathsfbi{N}_{0}}
\newcommand{\Pa}{\mathsfbi{P}_{\mathrm{a}}}
\newcommand{\Pz}{\mathsfbi{P}_{0}}
\newcommand{\Imat}{\mathsfbi{I}}
\newcommand{\Smat}{\mathsfbi{S}}
\newcommand{\Rev}{\mathsfbi{R}_{\mathrm{ev}}}
\newcommand{\Rcl}{\mathsfbi{R}_{\mathrm{cl}}}
\newcommand{\Rnull}{\mathsfbi{R}_{0}}
\newcommand{\Ractive}{\mathsfbi{R}_{\mathrm{a}}}

\begin{document}
	
	\maketitle
	
	\begin{abstract}
				Resolvent analysis treats turbulent fluctuations as the response of the mean flow to nonlinear forcing, and in its standard form selects the forcing by amplification alone. Whether turbulence follows this selection, and which part of its forcing can be modelled through the response, remains open. We address these questions by dividing the divergence-free forcing according to the pressure it produces, into a pressure-null sector that generates no pressure-gradient response and a pressure-active sector that complements the null sector. In parallel flows these sectors are the Squire and Orr--Sommerfeld forcing subspaces, respectively. Using direct numerical simulations of turbulent channel flow at $\Rey_\tau=180$, $550$ and $1000$, we show that, on average, nearly $90$ per cent of the forcing energy associated with the most energetic structures lies in the pressure-null sector, whereas the standard resolvent selects the pressure-active sector, where the amplification is several times stronger. In these structures, the pressure-active forcing supplies energy and generates streaks through lift-up, whereas the pressure-null forcing removes energy by opposing these streaks. These observations motivate the modelling assumption that the pressure-null forcing can be represented as feedback from the response in the form of an eddy viscosity, while the pressure-active forcing remains an input. This confines the eddy viscosity to the pressure-null sector, whereas in the conventional eddy-viscosity resolvent it acts on the whole response. When both are driven by the measured pressure-active forcing, the confined eddy viscosity matches the shape and the amplitude of the leading structures more closely.
	\end{abstract}

	\section{Introduction}
	\label{sec:introduction}

	Shear flows can strongly amplify disturbances even when all eigenmodes are asymptotically stable. This behaviour follows from the non-normality of the linearised Navier--Stokes operator and motivates an input--output description that quantifies how forcing is amplified into velocity fluctuations at each spatial scale and frequency \citep{TrefethenEtAl1993,JovanovicBamieh2005}. Resolvent analysis formalises this description. \citet{McKeonSharma2010} decompose the flow into a turbulent mean and fluctuations and treat the nonlinear fluctuation terms as forcing of the mean-flow linear operator. For each temporal frequency, and each wavenumber in the homogeneous directions of the mean flow, the resolvent maps that forcing to the velocity response. Its singular value decomposition ranks forcing and response pairs by linear amplification. The leading pair gives the largest response to unit-energy forcing under the chosen norms. This framework has been used to interpret coherent wall-turbulence structures \citep{McKeonSharma2010,SharmaMcKeon2013}, develop Reynolds-number scaling and self-similar descriptions \citep{MoarrefEtAl2013,KarbanEtAl2022}, describe coherent wavepackets in jets \citep{LesshafftEtAl2019,SchmidtEtAl2018}, characterise separated airfoil flows \citep{YehTaira2019}, analyse transonic buffet and high-speed flows \citep{KojimaEtAl2020,LiuEtAl2021,DwivediEtAl2019}, etc. Resolvent analysis has also been used for flow estimation \citep{IllingworthEtAl2018,TowneLozanoDuranYang2020,MartiniEtAl2020,AmaralEtAl2021} and control \citep{ToedtliEtAl2019,MartiniEtAl2022}.

	Amplification alone does not specify the forcing present in turbulence. In a turbulent flow the forcing is the nonlinear term itself, which redistributes energy among scales. Its net contribution to the volume-integrated fluctuation energy vanishes, but its work at a given scale can be positive or negative. \citet{SymonEtAl2021} found that nonlinear transfer removes energy from streamwise-constant streaks in turbulent channel flow and transfers it to other scales. A model driven only by independent white-in-time forcing supplies energy but does not represent this removal. Reproducing energetic coherent structures with resolvent analysis requires a forcing model that accounts for this `source--sink' separation.

	One established approach represents part of the nonlinear effect by adding an eddy viscosity to the linear operator \citep{ReynoldsHussain1972}, often with the profile of \citet{Cess1958}. The resulting energy sink improves the description of amplification and energetic structures in wall-bounded flows \citep{delAlamoJimenez2006,HwangCossu2010a,HwangCossu2010b,MorraEtAl2019,MadhusudananEtAl2019,SymonEtAl2023}. \citet{PickeringEtAl2021} optimised an eddy-viscosity distribution to improve agreement with spectral proper orthogonal decomposition (SPOD) modes in jets, and \citet{SymonEtAl2021} showed that replacing the molecular viscosity with a Cess profile approximates the energy removed from streamwise-constant channel-flow scales. With such a replacement, however, the eddy viscosity acts throughout the momentum equations, so it cannot distinguish forcing directions that supply energy from those that remove it.

	An alternative approach retains the mean-flow operator and represents the structure, or \emph{colour}, of the nonlinear forcing through its cross-spectral density. Direct numerical simulation (DNS) reveals strong spatial and componentwise forcing correlations, including substantial contributions along suboptimal resolvent directions \citep{MorraEtAl2021,NogueiraEtAl2021}. \citet{MorraEtAl2021} identified oblique streamwise vortices and a streamwise component in the solenoidal forcing whose responses oppose one another. Reproducing such opposition requires the relative phases of the forcing components as well as their energies. The cross-spectral density contains these phases, so a forcing model built from it can represent the opposition within the forcing, whereas an eddy viscosity represents it in the operator.

	The two approaches are connected through feedback, in which part of the forcing is written in terms of the response. When this part is modelled as a linear function of the velocity, it can be moved into the linear operator, leaving the rest of the forcing as the input. An eddy viscosity models this part as diffusion, with a viscosity that varies with wall-normal distance and is fitted to the mean flow. \citet{ZareEtAl2017} modelled the forcing as coloured in time, chosen to reproduce DNS velocity correlations, and showed that it is equivalent to white noise acting on a linear system modified by a feedback. Their feedback has no prescribed form and is fitted to the fluctuation statistics rather than to the mean flow. It can also be made low-rank, so that it enters the equations through a few forcing directions, whereas an eddy viscosity is full-rank, acting on every velocity component at every wall-normal position.

	This connection leaves a modelling choice open. It shows how the forcing modelled as response-dependent can be moved into the operator, but not which forcing contributions should be treated this way. The opposing components observed in DNS suggest that the forcing that adds energy to a coherent structure and the forcing that removes it may need different representations. The central question is therefore which part of the nonlinear forcing should remain an input and which part, if any, can be modelled as feedback from the response.

	The Orr--Sommerfeld/Squire (OS/Sq) decomposition \citep{SchmidHenningson2001} provides one such partition. In a parallel shear flow the linearised equations reduce to the Orr--Sommerfeld equation for the wall-normal velocity and the Squire equation for the wall-normal vorticity. The forcing defined in the standard resolvent contributes a term to each equation. Orr--Sommerfeld forcing drives wall-normal velocity, which the mean shear then converts into streaks, whereas Squire forcing acts on the wall-normal vorticity directly and produces no wall-normal velocity. \citet{RosenbergMcKeon2019} used this split to divide the resolvent modes into Orr--Sommerfeld and Squire families, and \citet{McMullenEtAl2020} fitted the weights of both families to DNS spectra of turbulent channel flow. They found the leading standard-resolvent modes nearly identical to the Orr--Sommerfeld family. When they split the modelled mean-square streamwise and spanwise velocities and the Reynolds shear stress into Orr--Sommerfeld, Squire and cross contributions, the cross term was negative in each. From this they inferred that the two responses are close to antiphase. \citet{PalyaEtAl2026} analyse a simplified resolvent that keeps only the path in which forcing generates wall-normal velocity, which then generates streaks through lift-up, and omits direct forcing of the Squire equation. They find that this omission changes the amplification of streamwise streaks little. These studies characterise the OS/Sq decomposition through the linear operator, with forcing that is optimal, white in time or implied by fitted mode weights.

	In this study we distinguish the two parts by the pressure they produce. In the linearised equations with divergence-free forcing, Squire forcing produces no wall-normal velocity \citep[][\S3.1]{SchmidHenningson2001} and therefore no pressure, whereas, for non-zero streamwise wavenumber, Orr--Sommerfeld forcing produces both. We define the partition from the pressure response of the primitive-variable resolvent, as the forcing that produces no pressure gradient and its orthogonal complement, and call these the pressure-null and pressure-active sectors. They coincide with the Squire and Orr--Sommerfeld forcing subspaces, respectively, but their definition does not require the Orr--Sommerfeld and Squire equations, which exist only for parallel flows. The partition therefore extends to non-parallel mean flows and complex geometries. To the best of our knowledge, the link between pressure and the Orr--Sommerfeld/Squire decomposition has not previously been used to examine how the nonlinear forcing divides between the two parts in the standard resolvent framework.

	Several studies change the directions selected by the resolvent by restricting the forcing to physically realisable disturbances \citep{KamalEtAl2023}, to actuator and sensor locations \citep{HerrmannEtAl2023}, or to forcing consistent with the structure of the nonlinear term \citep{LiuGayme2021}. We keep the full space of divergence-free forcing and ask a different question: how is the forcing in the turbulent flow divided between its pressure-null and pressure-active sectors? We extract the nonlinear forcing from DNS and measure the share of the forcing energy in each sector. We also investigate the interference between the responses the two sectors generate and the work each sector does on the response.

	We examine the partition both in the individual DNS realisations and in the forcing associated with the most energetic coherent structures. We identify the energetic structures by SPOD, which ranks flow structures at each frequency by their fluctuation energy \citep{Lumley1970,TowneSchmidtColonius2018} and is the usual reference for assessing resolvent responses \citep{MorraEtAl2019,AbreuEtAl2020,SchmidtEtAl2018,LesshafftEtAl2019,PickeringEtAl2020}. The forcing associated with each structure is obtained by resolvent-based extended proper orthogonal decomposition (RESPOD) \citep{KarbanEtAl2022,KarbanEtAl2023}, which builds on extended POD \citep{Boree2003} and extracts the part of the nonlinear forcing correlated with a chosen SPOD mode. This allows the forcing of an energetic structure to be compared directly with the forcing selected by amplification alone, that is, with the forcing modes of the resolvent operator.

	Finally, we use these measurements to build a model. The pressure-null forcing is modelled as an eddy viscosity acting on the response, with a wall-normal profile and an amplitude chosen from the measured pressure-null forcing of the leading mode. We first compare the leading response mode of the resulting eddy-viscosity resolvent with the leading SPOD mode. We then supply the measured pressure-active forcing as the input and compare this operator, in which the eddy viscosity acts on the whole response, with one in which the eddy-viscosity feedback is confined to the pressure-null sector, to determine where the feedback should act.

	The analysis covers turbulent channel flow at $\Rey_\tau\approx180$, $550$ and $1000$, for structures of aspect ratio $2$, $4$ and $8$ over a range of wavelengths. Section~\ref{sec:theory} sets out the OS/Sq forcing partition in \S\ref{sec:os_squire}, its pressure characterisation in \S\ref{sec:decomposition}, and the modal tools. Section~\ref{sec:database} describes the databases and spectral selection. Section~\ref{sec:sectors} compares how the forcing energy is distributed between the two sectors with the amplification available in each. Section~\ref{sec:source_sink_group} examines response interference, sector work and its dependence on streak energy. Section~\ref{sec:model} develops an eddy-viscosity model of the pressure-null forcing and tests it, and \S\ref{sec:conclusions} provides the concluding remarks.

	\section{Partition of the forcing and modal decompositions}
	\label{sec:theory}

	This section defines the two forcing sectors and the modal decompositions on which they are evaluated. For a parallel mean flow, the sectors can be defined in two equivalent ways, from the Orr--Sommerfeld and Squire equations and from the pressure response of the resolvent. We present both, because the first gives the physical interpretation of the sectors and the second extends this partition to flows in which these equations do not exist.

	\subsection{The Orr--Sommerfeld/Squire partition of the forcing}
	\label{sec:os_squire}

	We work with one Fourier mode of the flow at a time. Since the channel is homogeneous in the streamwise and spanwise directions, $x$ and $z$, and statistically stationary, we Fourier transform the forcing and the response in $x$, $z$ and time, and denote by a hat the coefficient at streamwise wavenumber $\alpha$, spanwise wavenumber $\beta$ and angular frequency $\omega$. We first set up coordinates in which the squared Euclidean norm of a forcing vector equals its energy. Forcing and response are both measured in the $L^2$ norm over the channel height, with the associated inner product,
\begin{equation}
\|\hat{\boldsymbol f}\|^2=\int_{-H}^{H}|\hat{\boldsymbol f}(y)|^2\,\mathrm dy,
\qquad
\langle\hat{\boldsymbol q},\hat{\boldsymbol f}\rangle=\int_{-H}^{H}\hat{\boldsymbol q}^{*}(y)\,\hat{\boldsymbol f}(y)\,\mathrm dy.
\label{eq:L2_norm}
\end{equation}
Here $\hat{\boldsymbol f}$ is the forcing, $\hat{\boldsymbol q}=[\hat u,\hat v,\hat w]^{\mathrm T}$ is the velocity response, $y$ is the wall-normal coordinate, $H$ is the channel half-height, with the walls at $y=\pm H$, and an asterisk denotes the complex conjugate transpose. For the velocity, the norm is proportional to the kinetic energy of the Fourier mode. Let $\Wmat$ be the diagonal matrix of quadrature weights for the three components, such that $\hat{\boldsymbol f}^{*}\Wmat\hat{\boldsymbol f}$ gives the discretised form of the norm in (\ref{eq:L2_norm}). We also restrict the forcing to its divergence-free, or solenoidal, part, because in incompressible flow the irrotational part of the forcing is balanced by pressure and produces no velocity (\S\ref{sec:source_sink}).

	For a fixed triplet $(\alpha,\beta,\omega)$, we expand the solenoidal forcing as
	\begin{equation}
		\hat{\boldsymbol f}_{\mathrm{s}}=\Bs\hat{\boldsymbol a},
		\label{eq:fs_coeff}
	\end{equation}
	where the columns of $\Bs$, constructed in appendix~\ref{app:sectors}, are divergence-free fields that vanish at the walls and are orthonormal in the norm (\ref{eq:L2_norm}). The solenoidal coordinates of a forcing field $\hat{\boldsymbol f}$ are
	\begin{equation}
\hat{\boldsymbol a}=\Bs^{\dagger}\hat{\boldsymbol f},
\qquad
\Bs^{\dagger}=\Bs^{*}\Wmat.
		\label{eq:Bs_dagger}
	\end{equation}
	With this choice $\Bs^{\dagger}\Bs=\Imat$, so that the Euclidean norm of the coefficients equals the norm (\ref{eq:L2_norm}) of the forcing field, $\|\hat{\boldsymbol a}\|=\|\hat{\boldsymbol f}_{\mathrm{s}}\|$. We write $\|\cdot\|$ for the Euclidean norm when it is applied to a coefficient vector and for the norm (\ref{eq:L2_norm}) when it is applied to a field. We define $k^{2}=\alpha^{2}+\beta^{2}$ and $\Delta=D^{2}-k^{2}$, with $D=\mathrm{d}/\mathrm{d}y$, and let the solenoidal forcing $\hat{\boldsymbol f}_{\mathrm{s}}=[\hat f_x,\hat f_y,\hat f_z]^{\mathrm{T}}$ satisfy $\ii\alpha\hat f_x+D\hat f_y+\ii\beta\hat f_z=0$. In the form of \citet{RosenbergMcKeon2019}, the forcing terms entering the Orr--Sommerfeld and Squire equations are
	\begin{equation}
		\hat g_{\mathrm{OS}}
		=
		-\ii\alpha D\hat f_x-k^{2}\hat f_y-\ii\beta D\hat f_z,
		\qquad
		\hat g_{\mathrm{Sq}}
		=
		\ii\beta\hat f_x-\ii\alpha\hat f_z.
		\label{eq:os_sq_forcing}
	\end{equation}
	Writing the first and third terms of $\hat g_{\mathrm{OS}}$ as $-D(\ii\alpha\hat f_x+\ii\beta\hat f_z)$ and using the solenoidal constraint gives $\hat g_{\mathrm{OS}}=\Delta\hat f_y$. The forcing vanishes at the walls, as the nonlinear term does under the no-slip condition, so $\Delta$ is invertible and the Orr--Sommerfeld forcing vanishes exactly when $\hat f_y$ does,
	\begin{equation}
		\hat g_{\mathrm{OS}}=0
		\quad\Longleftrightarrow\quad
		\hat f_y=0.
		\label{eq:gos_fy}
	\end{equation}

	The Squire forcing subspace $\mathcal F_{\mathrm{Sq}}$ is the set of solenoidal forcings that produce no Orr--Sommerfeld forcing, and the Orr--Sommerfeld forcing subspace $\mathcal F_{\mathrm{OS}}$ is its orthogonal complement in the norm (\ref{eq:L2_norm}). \citet{BuffatEtAl2011} use the same orthogonal splitting of a solenoidal field, into a part determined by its wall-normal component and a part with no wall-normal component, in the velocity decomposition of their spectral solver.

	Within the Squire subspace, the solenoidal constraint leaves a single free function. With $\hat f_y=0$, the constraint reduces to $\alpha\hat f_x+\beta\hat f_z=0$, so every Squire forcing takes the form
	\begin{equation}
		\hat{\boldsymbol f}_0(y)
		=
		\frac{1}{k}
		\begin{bmatrix}
			\beta\\
			0\\
			-\alpha
		\end{bmatrix}
		g(y),
		\label{eq:null_direction}
	\end{equation}
	with $g$ arbitrary. The direction $(\beta,0,-\alpha)/k$ is horizontal and perpendicular to the horizontal wavenumber vector $\boldsymbol k=(\alpha,0,\beta)^{\mathrm T}$. It therefore lies along the crests of the Fourier mode $\exp[\ii(\alpha x+\beta z-\omega t)]$, the lines of constant phase in the $x$--$z$ plane, which are inclined to the streamwise direction when $\alpha$ and $\beta$ are both non-zero. We use the orthonormal triad
	\begin{equation}
		\boldsymbol e_c=\frac{1}{k}\begin{bmatrix}\beta\\0\\-\alpha\end{bmatrix},
		\qquad
		\boldsymbol e_k=\frac{1}{k}\begin{bmatrix}\alpha\\0\\\beta\end{bmatrix},
		\qquad
		\boldsymbol e_y=\begin{bmatrix}0\\1\\0\end{bmatrix},
		\label{eq:crest_frame}
	\end{equation}
	in which $\boldsymbol e_c$ points along the crests, $\boldsymbol e_k=\boldsymbol k/k$ along the wavenumber vector and $\boldsymbol e_y$ normal to the walls. The velocity then has the components $\hat u_c=\boldsymbol e_c^{\mathrm T}\hat{\boldsymbol q}$, $\hat u_k=\boldsymbol e_k^{\mathrm T}\hat{\boldsymbol q}$ and $\hat v$, and the forcing has the components $\hat f_c$, $\hat f_k$ and $\hat f_y$, defined in the same way. Throughout this paper, the streak of a velocity or forcing field is defined as its crest component, $\hat u_c\boldsymbol e_c$ or $\hat f_c\boldsymbol e_c$, and the roll is its motion in the plane normal to the crests, $\hat u_k\boldsymbol e_k+\hat v\boldsymbol e_y$ or $\hat f_k\boldsymbol e_k+\hat f_y\boldsymbol e_y$. The roll axis therefore lies along the crests, and the mean shear turns the wall-normal velocity of the roll into streak velocity by lift-up. For streamwise-elongated structures, $\beta\gg\alpha$, the crests are nearly streamwise, $\boldsymbol e_c\to(1,0,0)^{\mathrm T}$ and $\boldsymbol e_k\to(0,0,1)^{\mathrm T}$, and the two definitions reduce to the usual ones, with the streak carried by the streamwise component and the roll by the cross-stream components. The Squire subspace has one degree of freedom per wall-normal node, the profile $g(y)$, and its orientation in the horizontal plane is set by the wavenumber pair alone. The pointwise projection of a forcing onto $\boldsymbol e_c$ is
	\begin{equation}
		\Smat
		=
		\boldsymbol e_c\boldsymbol e_c^{\mathrm T}
		=
		\frac{1}{k^{2}}
		\begin{bmatrix}
			\beta^{2} & 0 & -\alpha\beta\\
			0 & 0 & 0\\
			-\alpha\beta & 0 & \alpha^{2}
		\end{bmatrix},
		\label{eq:streak_projector}
	\end{equation}
	so that $\Smat\hat{\boldsymbol f}=\hat f_c\boldsymbol e_c$ at each $y$. This field has the form (\ref{eq:null_direction}), so $\Smat$ maps any forcing into $\mathcal F_{\mathrm{Sq}}$. The streak of a velocity or forcing field is obtained by applying $\Smat$ to it, and its roll by applying $\Imat-\Smat$. The two are perpendicular at every $y$ and hence orthogonal in the norm (\ref{eq:L2_norm}), so $\Smat$ is the orthogonal projector onto $\mathcal F_{\mathrm{Sq}}$. The orthogonal complement of $\mathcal F_{\mathrm{Sq}}$, the Orr--Sommerfeld subspace, consists of the solenoidal forcings that are pure roll,
	\begin{equation}
		\mathcal F_{\mathrm{OS}}
		=
		\{\hat{\boldsymbol f}_{\mathrm s}:\ (\Imat-\Smat)\hat{\boldsymbol f}_{\mathrm s}=\hat{\boldsymbol f}_{\mathrm s}\}
		=
		\{\hat{\boldsymbol f}_{\mathrm s}:\ \hat f_c=0\}.
		\label{eq:os_no_crest}
	\end{equation}

	Squire forcing does not drive the Orr--Sommerfeld equation, so its velocity response, $\hat{\boldsymbol q}_0=[\hat u_0,\hat v_0,\hat w_0]^{\mathrm T}$, has no wall-normal velocity. Continuity, $\ii\alpha\hat u_0+D\hat v_0+\ii\beta\hat w_0=\ii k\hat u_{k,0}+D\hat v_0=0$, then removes its component along $\boldsymbol e_k$ as well,
	\begin{equation}
		\hat v_0\equiv0,
		\qquad
		\hat u_{k,0}\equiv0.
		\label{eq:v0_zero}
	\end{equation}
	The response to Squire forcing is therefore a pure streak, $\hat{\boldsymbol q}_0=\hat u_{c,0}\boldsymbol e_c$, and all wall-normal velocity in the response comes from the Orr--Sommerfeld sector.

	\subsection{Pressure characterisation of the partition}
	\label{sec:decomposition}

	The two subspaces are separated by the pressure they generate, which gives the partition a second definition that does not refer to the Orr--Sommerfeld and Squire equations. Consider the primitive-variable resolvent relation
	\begin{equation}
		\begin{bmatrix}
			\hat{\boldsymbol q}\\
			\hat p
		\end{bmatrix}
		=
		\begin{bmatrix}
			\Hu\\
			\Hp
		\end{bmatrix}
		\hat{\boldsymbol f},
		\label{eq:primitive_resolvent}
	\end{equation}
	where $\hat p$ is the pressure response, and $\Hu$ and $\Hp$ are the parts of the resolvent operator that return the velocity and the pressure. For divergence-free forcing, the pressure is generated only by the mean shear acting on the wall-normal velocity. This is the rapid pressure of the rapid--slow decomposition \citep{Kim1989,LuharEtAl2014}. Using the pressure-Poisson equation, we show in appendix~\ref{app:sectors} that a Squire forcing, which produces no wall-normal velocity (\ref{eq:v0_zero}), generates no pressure-gradient response. For $\alpha\neq0$ the converse also holds, so that a forcing with no pressure-gradient response has no Orr--Sommerfeld forcing.

	The OS/Sq decomposition can therefore be constructed from the pressure response alone. The operator that maps solenoidal forcing coordinates to the pressure-gradient response is
	\begin{equation}
		\Js=\Gp\Hp\Bs,
		\label{eq:Js}
	\end{equation}
	where $\Gp=[\ii\alpha\Imat,\ \Dmat,\ \ii\beta\Imat]^{\mathrm T}$ maps the pressure to its three-component gradient, with $\Imat$ the identity matrix and $\Dmat$ the discrete form of $D$. The pressure-null forcing sector is the kernel of $\Js$, and the pressure-active forcing sector is its orthogonal complement,
	\begin{equation}
		\mathcal{N}_{0}=\ker(\Js),
		\qquad
		\mathcal{N}_{\mathrm{a}}=\mathcal{N}_{0}^{\perp}.
		\label{eq:nullspace}
	\end{equation}
	The singular value decomposition of $\Js$, detailed in appendix~\ref{app:sectors}, yields the matrices $\Nz$ and $\Na$, whose columns are the right singular vectors of $\Js$ with zero and non-zero singular values, respectively. These columns are orthonormal bases of $\mathcal N_0$ and $\mathcal N_{\mathrm a}$, and the corresponding orthogonal projectors are
	\begin{equation}
		\Pz=\Nz\Nz^{*},
		\qquad
		\Pa=\Na\Na^{*},
		\qquad
		\Pz+\Pa=\Imat.
		\label{eq:projectors}
	\end{equation}
	Because the projectors are orthogonal, the forcing energy divides between the sectors without a cross term,
	\begin{equation}
		\|\hat{\boldsymbol a}\|^2=\|\Pz\hat{\boldsymbol a}\|^2+\|\Pa\hat{\boldsymbol a}\|^2 .
		\label{eq:energy_split}
	\end{equation}
	For solenoidal forcing with $\alpha\neq0$ the two constructions give the same pair of subspaces,
	\begin{equation}
		\mathcal N_{0}=\mathcal F_{\mathrm{Sq}},
		\qquad
		\mathcal N_{\mathrm{a}}=\mathcal F_{\mathrm{OS}}, \label{eq:os_equivalence}
	\end{equation}
	and the pressure-null projector in forcing coordinates is the pointwise streak projector of (\ref{eq:streak_projector}),
	\begin{equation}
		\Bs\Pz\Bs^{\dagger}=\Smat.
		\label{eq:null_is_streak}
	\end{equation}
	The construction of $\mathcal N_0$ uses only the primitive-variable resolvent (\ref{eq:primitive_resolvent}), the gradient operator $\Gp$ and a basis of solenoidal forcing, all of which can be formed for any mean flow. The Orr--Sommerfeld and Squire equations, in contrast, exist only for a parallel mean flow $U(y)$, and the equivalence (\ref{eq:os_equivalence}) is needed only to interpret the sectors in that case. The pressure definition therefore also applies to global resolvent operators of non-parallel mean flows, where no Orr--Sommerfeld/Squire split is available. In such flows the pressure-null forcing still produces no pressure-gradient response, but the properties derived in \S\ref{sec:os_squire}, such as the absence of wall-normal velocity in its response, need not hold.	Because the pressure-based definition is more general, we name the sectors pressure-null and pressure-active, and use these names for the Squire and Orr--Sommerfeld subspaces of the channel as well.

	\subsection{The standard resolvent and its restriction to the pressure-null and pressure-active sectors}
	\label{sec:restricted}

	The velocity resolvent in the same solenoidal forcing coordinates is
	\begin{equation}
		\Rs
		=
		\Hu\Bs,
		\label{eq:Rs}
	\end{equation}
	so that $\hat{\boldsymbol q}=\Rs\hat{\boldsymbol a}$ is the velocity response to the solenoidal forcing $\Bs\hat{\boldsymbol a}$. Conventional resolvent analysis solves
	\begin{equation}
		\max_{\hat{\boldsymbol a}\neq\boldsymbol{0}}
		\frac{\|\Rs\hat{\boldsymbol a}\|^{2}}{\|\hat{\boldsymbol a}\|^{2}}
		\label{eq:standard_optimisation}
	\end{equation}
	to identify the forcing direction that produces the largest response energy per unit forcing energy. The singular values and singular vectors of $\Rs$, and of the other operators that map forcing coordinates to the response, are taken with respect to these two norms throughout. A singular value is then the square root of an energy gain, and the left singular vectors are orthonormal in the norm (\ref{eq:L2_norm}). Restricting the forcing to one sector at a time, with the orthonormal bases $\Nz$ and $\Na$, gives the restricted resolvents
	\begin{equation}
		\Rnull=\Rs\Nz,
		\qquad
		\Ractive=\Rs\Na.
		\label{eq:restricted_resolvents}
	\end{equation}
	Their leading singular values satisfy
	\begin{align}
		\sigma_{1,0}
		&=
		\max_{\substack{\hat{\boldsymbol a}\in\mathcal{N}_{0}\\
				\|\hat{\boldsymbol a}\|=1}}
		\|\Rs\hat{\boldsymbol a}\|,
		\\
		\sigma_{1,\mathrm{a}}
		&=
		\max_{\substack{\hat{\boldsymbol a}\in\mathcal{N}_{\mathrm{a}}\\
				\|\hat{\boldsymbol a}\|=1}}
		\|\Rs\hat{\boldsymbol a}\|.
	\end{align}
	The ratio $\sigma_{1,\mathrm{a}}/\sigma_{1,0}$ compares the largest amplification available in each sector at equal forcing energy.

	The restricted resolvents are the Orr--Sommerfeld and Squire families of \citet{RosenbergMcKeon2019}, written in primitive variables. \citet{RosenbergMcKeon2019} measure forcing and response with a kinetic-energy norm in the wall-normal velocity and vorticity, which for divergence-free fields equals the norm (\ref{eq:L2_norm}), and each sector drives only one of the two equations. A pressure-active forcing lies in $\mathcal N_{\mathrm a}=\mathcal F_{\mathrm{OS}}$ by (\ref{eq:os_equivalence}), so by (\ref{eq:os_no_crest}) it has no streak and thus no Squire forcing. Similarly, a pressure-null forcing has no wall-normal component, so its Orr--Sommerfeld forcing vanishes by (\ref{eq:gos_fy}).

	\subsection{SPOD and RESPOD modes}
	\label{sec:spod_respod}

	The decomposition described above applies to any forcing vector. We evaluate it in this study on the forcing selected by the standard resolvent, given by the right singular vectors of $\Rs$, and on the nonlinear forcing that generates the most energetic structures of the flow, obtained from the DNS by RESPOD.

	Let $\hat{\boldsymbol{q}}^{(k)}$ and $\hat{\boldsymbol{a}}^{(k)}$, $k=1,\dots,N$, denote Fourier realisations of the response and of the solenoidal forcing coefficients at a given $(\alpha,\beta,\omega)$. SPOD \citep{TowneSchmidtColonius2018} decomposes the response realisations into modes $\hat{\boldsymbol\psi}_j$ that are orthonormal in the $L^2$ norm and ranked by the mean fluctuation energy $\lambda_1\geq\lambda_2\geq\cdots$ they capture. The mode index $j$ is the mode number used throughout the results. RESPOD \citep{KarbanEtAl2022} is the extended POD \citep{Boree2003} of the forcing with respect to these modes, and gives a forcing mode $\hat{\boldsymbol a}_j$ for each $\hat{\boldsymbol\psi}_j$. We compute both sets of modes from the realisations as described in appendix~\ref{app:modes}. Every realisation satisfies $\hat{\boldsymbol q}^{(k)}=\Rs\hat{\boldsymbol a}^{(k)}$, and the modes inherit this relation,
	\begin{equation}
		\Rs\hat{\boldsymbol a}_j=\hat{\boldsymbol\psi}_j,
		\label{eq:respod_closure}
	\end{equation}
	so the response to each sector's part of $\hat{\boldsymbol a}_j$ can be computed with $\Rs$.

	\section{Flow configuration and numerical databases}
	\label{sec:database}

	We use DNS databases of incompressible turbulent channel flow at three friction Reynolds numbers, each analysed at the same set of streamwise-to-spanwise aspect ratios. The flow is bounded by two parallel no-slip walls separated by a distance $2H$. The friction Reynolds number is defined as
	\begin{equation}
		\Rey_\tau=\frac{u_\tau H}{\nu},
	\end{equation}
	where $u_\tau$ is the friction velocity and $\nu$ is the kinematic viscosity. A superscript $+$ denotes a quantity in wall units, in which lengths are scaled by $\nu/u_\tau$, velocities by $u_\tau$ and times by $\nu/u_\tau^2$. We consider three databases, with actual friction Reynolds numbers $\Rey_\tau=179$, $543$ and $996$. For convenience, we refer to these cases throughout the paper by their nominal values $\Rey_\tau=180$, $550$ and $1000$, respectively.

	The DNS were obtained using the ChannelFlow pseudo-spectral solver \citep{GibsonEtAl2019}. The streamwise and spanwise directions are periodic and are discretised using Fourier expansions, Chebyshev polynomials are used in the wall-normal direction, and $3/2$ dealiasing is applied in the homogeneous directions. The $\Rey_\tau=180$ case uses a domain $L_x\times L_y\times L_z=4\pi H\times2H\times2\pi H$, whereas the two higher-Reynolds-number cases use the more compact domain $2\pi H\times2H\times\pi H$. Table~\ref{tab:dns_parameters} summarises the principal numerical parameters of the three databases, together with a second Reynolds number, $\Rey=U_bH/\nu$, based on the bulk velocity $U_b$. Initial transients were discarded before the snapshots retained for statistical analysis were collected \citep{AmaralEtAl2021}.

	\begin{table*}
		\centering
		\small
		\setlength{\tabcolsep}{5pt}
		\begin{tabular}{ccccccccc}
			\hline
			$\Rey_\tau$ &
			$\Rey$ &
			$(L_x,L_y,L_z)/H$ &
			$N_x$ &
			$N_y$ &
			$N_z$ &
			$\Delta x^+$ &
			$\Delta z^+$ &
			$(\Delta y^+_{\min},\Delta y^+_{\max})$ \\
			\hline
			179 &
			2800 &
			$(4\pi,\,2,\,2\pi)$ &
			192 & 129 & 192 &
			11.71 & 5.86 &
			$(0.0539,\,4.39)$
			\\
			543 &
			10000 &
			$(2\pi,\,2,\,\pi)$ &
			384 & 257 & 384 &
			8.89 & 4.44 &
			$(0.0409,\,6.66)$
			\\
			996 &
			20000 &
			$(2\pi,\,2,\,\pi)$ &
			484 & 385 & 484 &
			12.93 & 6.47 &
			$(0.0333,\,8.15)$
			\\
			\hline
		\end{tabular}
		\caption{Parameters of the turbulent-channel DNS databases used in the present analysis. $\Rey=U_bH/\nu$, and $N_x$, $N_y$ and $N_z$ denote the spatial resolution. The Fourier discretisation uses $3/2$ dealiasing in the homogeneous directions. The friction Reynolds number is obtained from the wall shear of the mean velocity profile and the spacings in wall units from the grid of each database; the sampling of the time histories is given in table~\ref{tab:welch}.}
		\label{tab:dns_parameters}
	\end{table*}

	The $\Rey_\tau\approx180$ database was used by \citet{MartiniEtAl2020} for resolvent-based estimation, and the $\Rey_\tau\approx180$ and $550$ databases by \citet{MorraEtAl2021} for nonlinear forcing statistics. The $\Rey_\tau\approx550$ database was used by \citet{KarbanEtAl2022} for RESPOD, who also checked its statistical convergence through the mean-momentum balance. \citet{AmaralEtAl2021} used all three databases to estimate the velocity and pressure fluctuations from wall measurements of shear stress and pressure. The mean and fluctuation profiles agree with the DNS of \citet{delAlamoJimenez2003} at the two lower Reynolds numbers and with that of \citet{LeeMoser2015} at the highest.

	The present work uses the same velocity, pressure and nonlinear-forcing data. For each Reynolds number, we consider three aspect ratios of the streamwise wavelength, $\lambda_x=2\pi/\alpha$, to the spanwise wavelength, $\lambda_z=2\pi/\beta$,
	\begin{equation}
		AR=\frac{\lambda_x}{\lambda_z},
	\end{equation}
	with $AR=2$, $4$ and $8$. The resulting nine $(\Rey_\tau,AR)$ combinations allow the effects of aspect ratio and Reynolds number to be assessed separately. Below, we refer to a wavenumber pair $(\alpha,\beta)$ as a wave.

	In each database the longest stored wave, the streamwise fundamental, has $\lambda_x$ equal to the streamwise period, $L_x=4\pi H$ at $\Rey_\tau\approx180$ and $2\pi H$ at the other two. With a single wavelength in the domain, the fundamental wave is the most constrained by the imposed periodicity. We therefore exclude it from all analyses, which removes one wave per $(\Rey_\tau,AR)$ combination. The effect of this exclusion is noted in \S\ref{sec:source_sink}. The next longest wave, $\lambda_x=L_x/2$, is retained.

	Overall, we use $83$ waves. Table~\ref{tab:welch} gives their number and the range of streamwise wavelengths at each Reynolds number. The number of available waves can depend on $AR$, because pairs that cannot be represented adequately on the corresponding Fourier grid are excluded. Reynolds-number comparisons are therefore made over the waves available at each $AR$, without a one-to-one match of wavelengths across Reynolds numbers.

	We divide each time history into blocks of $N_{\mathrm{fft}}$ samples that overlap by $80$ per cent. Each block is multiplied by the window of \citet{MartiniEtAl2019arXiv} and Fourier transformed. As detailed in appendix~\ref{app:modes}, we also add the forcing-like term that this window introduces, so that the resolvent relates the windowed forcing and response exactly. Each transformed block constitutes one Fourier realisation, and averaging over the blocks gives the cross-spectral densities, as in Welch's method. Table~\ref{tab:welch} gives the sampling interval, duration, $N_{\mathrm{fft}}$, block length and number of blocks for each database.

	The databases at $\Rey_\tau\approx550$ and $1000$ contain fewer time samples than that at $180$, so the same $N_{\mathrm{fft}}$ yields fewer blocks and hence fewer Fourier realisations per wave. To compensate for this reduction, we double the ensemble at these two Reynolds numbers using the mirrored wavenumber pair $(-\alpha,\beta)$. Since the velocity is real, the complex conjugate of $(-\alpha,\beta)$ is $(\alpha,-\beta)$. Reversing its spanwise velocity and forcing components maps it onto $(\alpha,\beta)$ under the spanwise reflection symmetry of the channel. Therefore, it has the same statistics as the $(\alpha,\beta)$ history, and spanwise homogeneity makes the two uncorrelated. The mapped history therefore provides a second set of realisations of the same wave, which we pool with those of $(\alpha,\beta)$ into one ensemble. Pooling doubles the number of realisations per wave at these two Reynolds numbers, to the values given in table~\ref{tab:welch}. At $\Rey_\tau\approx180$ no pooling is applied.

	\begin{table}
		\centering
		\begin{tabular}{lcccccccc}
			\toprule
			$\Rey_\tau$ & $\Delta t\,U_b/H$ & samples & $T\,U_b/H$ & $N_{\mathrm{fft}}$
			& block length $U_b/H$ & realisations per wave & waves & $\lambda_x^+$\\
			\midrule
			$180$  & $0.5$  & $7801$ & $3900$  & $512$ & $256$  & $72$ & $18$ & $321$--$1124$\\
			$550$  & $0.1$  & $3000$ & $299.9$ & $512$ & $51.2$ & $50$ & $30$ & $262$--$1706$\\
			$1000$ & $0.05$ & $5241$ & $262$   & $512$ & $25.6$ & $94$ & $35$ & $447$--$3128$\\
			\bottomrule
		\end{tabular}
		\caption{Welch parameters of the three databases and the waves analysed. $\Delta t$ is the sampling interval, $T$ the duration and $N_{\mathrm{fft}}$ the number of samples per block; consecutive blocks overlap by $80$ per cent. The last two columns give the number of waves retained and the range of their streamwise wavelengths.}
		\label{tab:welch}
	\end{table}

\begin{figure}
		\centering
		\includegraphics[width=.66\textwidth]{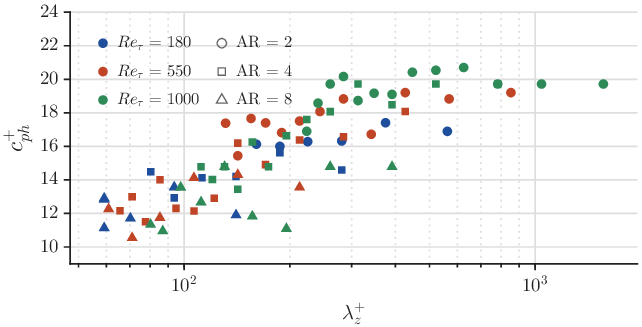}
		\caption{Phase speed at the selected frequency in wall units, $c_{\mathrm{ph}}^+$, against $\lambda_z^+$, one marker per wave. Colours denote $\Rey_\tau$ (blue, $180$; red, $550$; green, $1000$) and markers $AR$ (circles, $2$; squares, $4$; triangles, $8$).}
		\label{fig:phase_speed}
	\end{figure}

	For every wavenumber pair, we select the temporal frequency $\omega_{\max}$ at the peak of the power spectral density of the response, summed over the three velocity components and integrated over the channel height. The same criterion is used for all Reynolds numbers and aspect ratios, and all quantities reported below are evaluated at $\omega_{\max}$ unless stated otherwise. Figure~\ref{fig:phase_speed} shows the selected frequencies in terms of the phase speed of each wave, $c_{\mathrm{ph}}=\omega_{\max}/\alpha$, in wall units. It is seen that $c_{\mathrm{ph}}^+$ lies between about $10$ and $21$ and increases with the spanwise wavelength. At a given $\lambda_z^+$, the waves of lower aspect ratio generally travel faster.

	\section{Forcing energy and amplification in the two sectors}
	\label{sec:sectors}

	We have shown that the energy of any solenoidal forcing splits into a pressure-null and a pressure-active part that add up, and that each part has its own restricted resolvent. This section measures how the two parts are shared in the nonlinear forcing and in the forcing selected by the standard resolvent (\S\ref{sec:occupation}), and then compares the gains of the two restricted resolvents (\S\ref{sec:amplification}).

	\subsection{Distribution of the forcing energy between the sectors}
	\label{sec:occupation}

	We start the analysis by decomposing the response at each $(\alpha,\beta,\omega)$ into its SPOD modes $\hat{\boldsymbol\psi}_j$ and the nonlinear forcing into the corresponding RESPOD modes $\hat{\boldsymbol a}_j$. To measure how a forcing vector $\hat{\boldsymbol a}$ occupies the two sectors, we use its pressure-null fraction, the share of its solenoidal forcing energy in the pressure-null sector,
	\begin{equation}
		\rho_{0}(\hat{\boldsymbol a})
		=
		\frac{\|\Nz^{*}\hat{\boldsymbol a}\|^2}
		{\|\hat{\boldsymbol a}\|^2}.
		\label{eq:null_fraction}
	\end{equation}
	By (\ref{eq:energy_split}), the pressure-active sector holds the remaining share, $1-\rho_0$. We evaluate the fraction for the RESPOD modes $\hat{\boldsymbol a}_j$ and for the right singular vectors of $\Rs$, which are the forcing directions selected by the standard resolvent.

	\begin{figure}
		\centering
		\includegraphics[width=\textwidth]{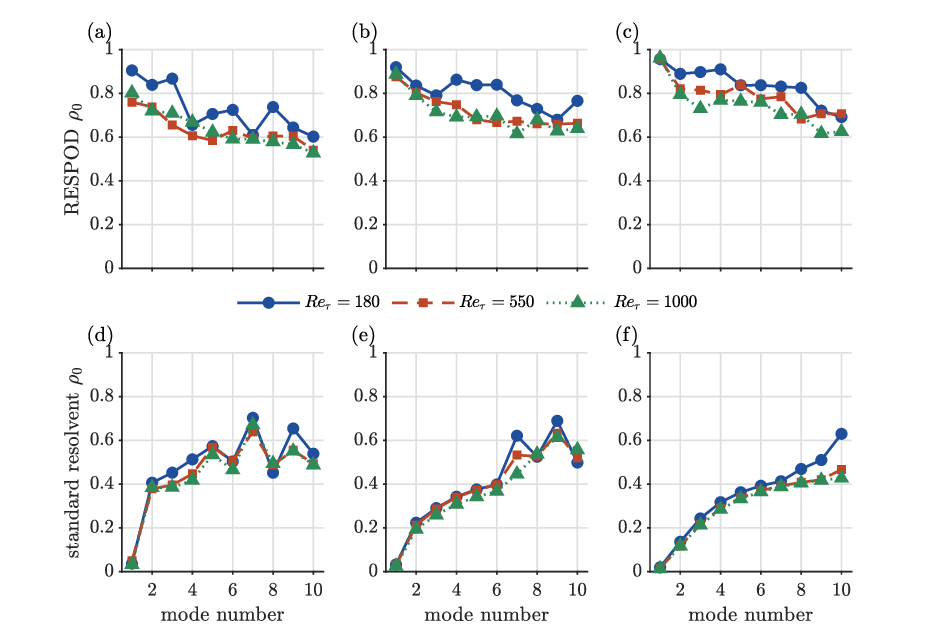}
		\caption{Pressure-null fraction $\rho_0$ of (\ref{eq:null_fraction}) against mode number. Top row (a--c): RESPOD forcing. Bottom row (d--f): standard-resolvent forcing. Columns: (a,d) $AR=2$, (b,e) $AR=4$, (c,f) $AR=8$. Blue solid lines with circles, $\Rey_\tau\approx180$; red dashed lines with squares, $550$; green dotted lines with triangles, $1000$. Each curve is the mean over the waves retained in that combination.}
		\label{fig:sector_occupation}
	\end{figure}

	Figure~\ref{fig:sector_occupation} shows the pressure-null fraction $\rho_0$ of (\ref{eq:null_fraction}) against mode number, and table~\ref{tab:occupation} gives its value at the leading mode averaged over the waves of each combination. The nonlinear forcing associated with the most energetic structures is predominantly pressure-null, the leading RESPOD forcing being between $76$ and $96$ per cent pressure-null at every Reynolds number and aspect ratio. The forcing selected by the standard resolvent occupies the opposite sector, being only $1$ to $5$ per cent pressure-null. This preference of the standard resolvent agrees with \citet{McMullenEtAl2020}, who found the leading standard-resolvent modes nearly identical to those of the Orr--Sommerfeld family \citep[see also][]{DawsonMcKeon2019}. A representation fitted to response spectra does not, however, determine how the nonlinear forcing itself is distributed between the sectors, which we measure here from the forcing data.
	\begin{table}
		\centering
		\begin{tabular}{lcccccc}
			\toprule
			& \multicolumn{3}{c}{RESPOD} & \multicolumn{3}{c}{standard resolvent}\\
			\cmidrule(lr){2-4}\cmidrule(lr){5-7}
			$\Rey_\tau$ & $AR=2$ & $AR=4$ & $AR=8$ & $AR=2$ & $AR=4$ & $AR=8$\\
			\midrule
			$180$  & $0.90$ & $0.92$ & $0.96$ & $0.04$ & $0.03$ & $0.02$\\
			$550$  & $0.76$ & $0.88$ & $0.96$ & $0.05$ & $0.03$ & $0.02$\\
			$1000$ & $0.80$ & $0.89$ & $0.96$ & $0.03$ & $0.02$ & $0.01$\\
			\bottomrule
		\end{tabular}
		\caption{Pressure-null fraction $\rho_0$ of the leading forcing mode, averaged over the waves of each combination.}
		\label{tab:occupation}
	\end{table}
	The separation is confined to the leading modes. By mode ten the ranges of the two fractions overlap (figure~\ref{fig:sector_occupation}). The results below therefore concern the energetically dominant structures and not the forcing space as a whole.

	These fractions measure forcing energy, and the contribution of each sector to the response can differ.
	\begin{figure}
		\centering
		\includegraphics[width=\textwidth]{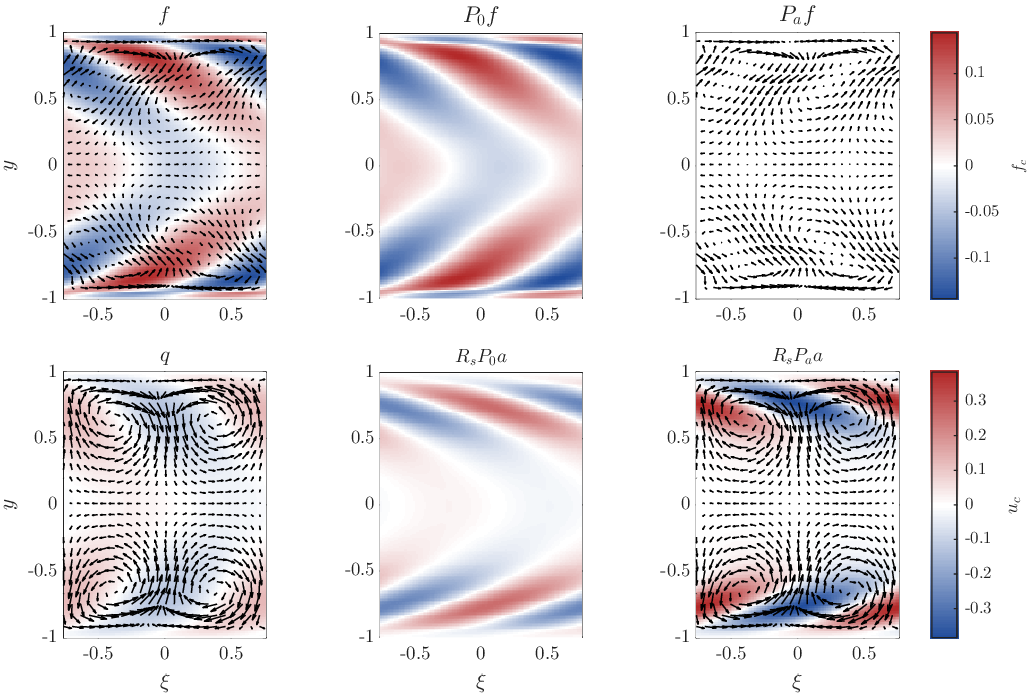}
		\caption{Plane normal to the wave crests at $\Rey_\tau\approx180$, $AR=4$, $\lambda_x^+=1124$, $\lambda_z^+=281$, with $\xi$ the coordinate along $\boldsymbol e_k$ of (\ref{eq:crest_frame}). Top row: the RESPOD forcing and its pressure-null and pressure-active parts. Bottom row: the leading SPOD response and the responses generated by each part. Colour is the component along $\boldsymbol e_c$, and the arrows are the in-plane vectors formed by the components along $\boldsymbol e_k$ and $\boldsymbol e_y$; the colour and arrow scales are common within each row. The absolute phase of the mode is arbitrary, whereas the relative phase between panels is determined by the flow data.}
		\label{fig:mode_reconstruction}
	\end{figure}
	Figure~\ref{fig:mode_reconstruction} shows one case, at $\lambda_x^+=1124$ and $AR=4$, in the plane normal to the wave crests, which are inclined at $14^\circ$ to the streamwise direction. The coordinate $\xi$ runs along $\boldsymbol e_k$, and the mode depends on $x$ and $z$ only through $\alpha x+\beta z=k\xi$, so this plane contains the whole mode. Colour shows the streak, through $\hat f_c$ for the forcing and $\hat u_c$ for the velocity. The streak lies along the pressure-null direction of (\ref{eq:null_direction}). The arrows show the two components in the plane. The RESPOD forcing contains both a streak and a roll, and the sector decomposition separates the two exactly. The pressure-null part appears in colour only, because it is directed along $\boldsymbol e_c$. The pressure-active part is the roll and appears as arrows only, because it has no streak (\ref{eq:os_no_crest}).

	This exact separation follows from the equivalence (\ref{eq:os_equivalence}), which we verified on the discretised operators (not shown here). In every wavenumber pair, the kernel of $\Js$ reproduces the Squire subspace, and its orthogonal complement the Orr--Sommerfeld subspace, to within numerical round-off.

	In the response, the separation holds only partly. The null forcing generates a pure streak by (\ref{eq:v0_zero}), which also appears in colour only. The active forcing generates a roll, shown by the arrows, together with the streak that the roll drives by lift-up, shown in colour. Although it carries only a small fraction of the forcing energy, its response is of comparable magnitude to the full one.

	\subsection{Amplification of the two sectors}
	\label{sec:amplification}

	Figure~\ref{fig:mode_reconstruction} shows that the pressure-null dominance of the forcing does not extend to the response. Since $\Rs\hat{\boldsymbol a}_j=\hat{\boldsymbol\psi}_j$, a sector that $\Rs$ amplifies weakly can hold much of the energy of $\hat{\boldsymbol a}_j$ while contributing little to $\hat{\boldsymbol\psi}_j$. To measure how differently the operator amplifies the two sectors, we compare the leading gains of the restricted resolvents (\ref{eq:restricted_resolvents}).

	\begin{figure}
		\centering
		\includegraphics[width=\textwidth]{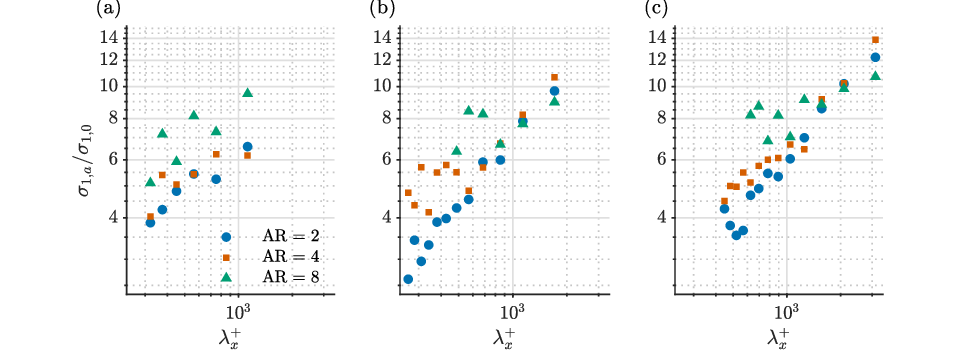}
		\caption{Ratio of the leading gains of the restricted resolvents, $\sigma_{1,\mathrm{a}}/\sigma_{1,0}$, against streamwise wavelength, at (a) $\Rey_\tau\approx180$, (b) $550$ and (c) $1000$. One marker per wave: blue circles $AR=2$, orange squares $AR=4$, green triangles $AR=8$. Both axes logarithmic.}
		\label{fig:gain_separation}
	\end{figure}

	Figure~\ref{fig:gain_separation} shows the ratio $\sigma_{1,\mathrm{a}}/\sigma_{1,0}$ for every wave. It is seen that the active sector has the larger leading gain in every wave, with the ratio ranging from $2.6$ to $13.9$. The ratio increases with $\lambda_x^+$, most steeply at $AR=2$, and is generally larger at higher aspect ratio. Since the active sector has the larger gain, an optimisation that ranks forcing directions by response energy alone favours it. This is consistent with the standard-resolvent forcing modes of table~\ref{tab:occupation}, which lie almost entirely in the active sector. In an additional analysis not shown here, we compared $\sigma_{1,\mathrm{a}}$ with the leading gain $\sigma_1$ of $\Rs$ and observed that the two differ by less than $6$ per cent in every wave. \citet{PalyaEtAl2026} similarly found that omitting the direct forcing of the Squire equation changes the gain of streamwise streaks little.

	\begin{figure}
		\centering
		\includegraphics[width=\textwidth]{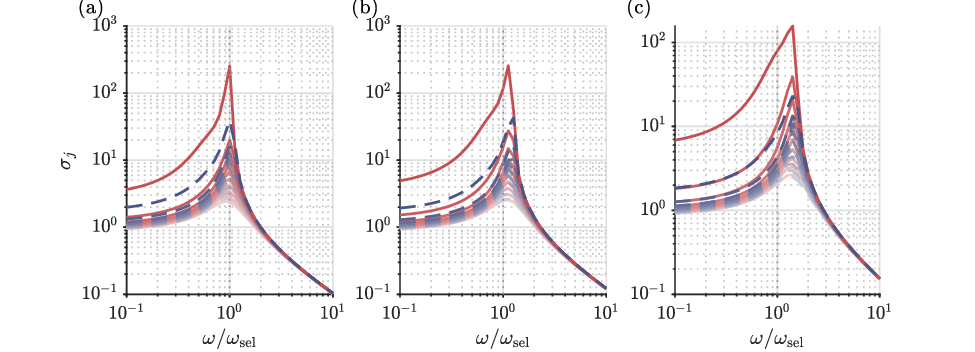}
		\caption{Singular values of the restricted resolvents against frequency, normalised by the selected frequency, for one wave at $\Rey_\tau\approx180$ and $\lambda_x^+=1124$, with (a) $AR=2$, (b) $AR=4$ and (c) $AR=8$. Red solid curves, pressure-active sector; blue dashed curves, pressure-null sector. Within each sector the darkest curve is the leading mode and the shade lightens with mode index up to ten. The vertical dotted line marks $\omega=\omega_{\max}$. Both axes logarithmic.}
		\label{fig:gain_spectra}
	\end{figure}

	The ratios in figure~\ref{fig:gain_separation} are evaluated at the selected frequency $\omega_{\max}$ of each wave. At other frequencies the ratio changes, although the active and null sectors themselves do not depend on frequency. The operator $\Js$ depends on $\omega$ through $\Hp$, but its kernel is the Squire subspace (\ref{eq:os_equivalence}), and the pressure-null projector is the streak projector $\Smat$ (\ref{eq:null_is_streak}), which is built from $\alpha$ and $\beta$ alone. Figure~\ref{fig:gain_spectra} shows the leading singular values of $\Ractive$ and $\Rnull$ over two decades in frequency. Their ratio is largest near $\omega_{\max}$ and is practically unity at $10\,\omega_{\max}$ at every Reynolds number and aspect ratio. Near $\omega_{\max}$ the difference between the two gains comes from lift-up. A forcing with a wall-normal component drives a roll, whose wall-normal velocity generates the streak through the mean shear. Such forcing lies in the active sector. The response to a pressure-null forcing has no wall-normal velocity. Its streak is forced directly, which is not as efficient, and thus, yields lower gain. For $\omega\gg\omega_{\max}$, $\Rs\rightarrow(-\ii\omega)^{-1}\Bs$, which amplifies every direction equally. Thus, the leading gains of the active and null sectors collapse at high frequencies, as seen in figure~\ref{fig:gain_spectra}.

	\section{Source and sink roles of the two sectors}
	\label{sec:source_sink_group}

	The response modes in figure~\ref{fig:mode_reconstruction} show that, for this wave, the streak generated by the pressure-active forcing is in phase with the streak of the full response, whereas the streak generated by the pressure-null forcing opposes it. In this section we test whether this holds for every wave, and whether the active and null sectors act as a source and a sink of energy.

	\subsection{Interference between the sector responses}
	\label{sec:interference}

	We first examine the phase between the responses of the two sectors. Writing $\hat{\boldsymbol a}=\Pz\hat{\boldsymbol a}+\Pa\hat{\boldsymbol a}$ and applying $\Rs$ gives the two sector responses,
	\begin{equation}
		\hat{\boldsymbol q}=\hat{\boldsymbol q}_{0}+\hat{\boldsymbol q}_{\mathrm{a}},
		\qquad
		\hat{\boldsymbol q}_{0}=\Rs\Pz\hat{\boldsymbol a},
		\qquad
		\hat{\boldsymbol q}_{\mathrm{a}}=\Rs\Pa\hat{\boldsymbol a}.
		\label{eq:sector_responses}
	\end{equation}
	By (\ref{eq:v0_zero}) and continuity, $\hat{\boldsymbol q}_{0}$ is a pure streak, whereas $\hat{\boldsymbol q}_{\mathrm{a}}$ carries both a streak and a roll. We make the comparison on the streak parts alone, with $\hat{\boldsymbol q}_{\mathrm{a,s}}=\Smat\hat{\boldsymbol q}_{\mathrm{a}}$, since the roll is orthogonal to $\hat{\boldsymbol q}_{0}$ and cannot interfere with it.

	The streak part of the response is $\hat{\boldsymbol q}_{\mathrm{s}}=\hat{\boldsymbol q}_{0}+\hat{\boldsymbol q}_{\mathrm{a,s}}$, whose energy is
	\begin{equation}
		\|\hat{\boldsymbol q}_{\mathrm{s}}\|^{2}=\bigl(\|\hat{\boldsymbol q}_{0}\|^{2}+\|\hat{\boldsymbol q}_{\mathrm{a,s}}\|^{2}\bigr)(1+\chi_{\mathrm{s}}),
		\qquad
		\chi_{\mathrm{s}}
		=
		\frac{2\,\mathrm{Re}\langle \hat{\boldsymbol q}_{0},\hat{\boldsymbol q}_{\mathrm{a,s}}\rangle}
		{\|\hat{\boldsymbol q}_{0}\|^{2}+\|\hat{\boldsymbol q}_{\mathrm{a,s}}\|^{2}}.
		\label{eq:chi_s}
	\end{equation}
	The interference coefficient $\chi_{\mathrm{s}}$ is the fractional change in streak energy relative to the two sectors acting independently. It lies between $-1$ and $+1$, indicating a complete cancellation or perfect alignment, respectively.

\begin{figure}
		\centering
		\includegraphics[width=.9\textwidth]{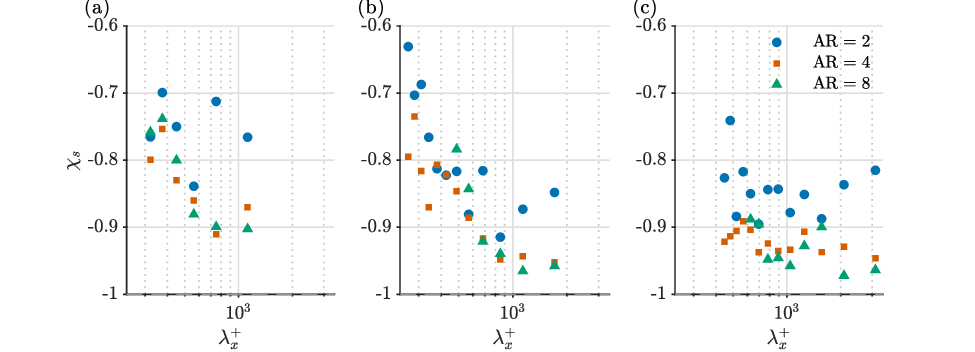}
		\caption{Interference coefficient $\chi_{\mathrm{s}}$ of (\ref{eq:chi_s}) between the pressure-null response and the streak part of the pressure-active response, for the leading mode, against streamwise wavelength, at (a) $\Rey_\tau\approx180$, (b) $550$ and (c) $1000$. One marker per wave: blue circles $AR=2$, orange squares $AR=4$, green triangles $AR=8$. The horizontal dashed line at $-1$ is complete cancellation.}
		\label{fig:sector_phase}
	\end{figure}

	Figure~\ref{fig:sector_phase} shows $\chi_{\mathrm{s}}$ for the leading mode.
	The null response opposes the active one in every wave. The mean of $\chi_{\mathrm{s}}$ over each $(\Rey_\tau,AR)$ combination lies between $-0.76$ and $-0.93$. At $\Rey_\tau\approx180$ and $550$ the cancellation strengthens with $\lambda_x^+$, whereas at $\Rey_\tau\approx1000$ the trend is weak compared with the scatter. At a given $\lambda_x^+$, $\chi_{\mathrm{s}}$ is nearly the same at $AR=4$ and $AR=8$, while at $AR=2$ it is slightly less negative. Plotted against $\lambda_z^+$ (not shown), the three aspect ratios separate, and at equal $\lambda_z^+$ the longer waves cancel more. The cancellation therefore depends on the streamwise length of the structure more than on its spanwise width.

	Several earlier studies report a similar opposition. \citet{McMullenEtAl2020} fitted a resolvent representation to DNS spectra and found the cross term between the Orr--Sommerfeld and Squire families negative in the variances of the streamwise and spanwise velocity and in the Reynolds shear stress. They interpret this as the two families lying close to antiphase. \citet{Rosenberg2018} found a similar opposition in the wall-normal vorticity of DNS at $\Rey_\tau\approx185$. \citet{MorraEtAl2021} showed that, in the solenoidal forcing at $\Rey_\tau\approx180$ and $550$, the response to the wall-normal and spanwise components, shaped as quasi-streamwise vortices, is counteracted by the response to the streamwise component. For streamwise-elongated structures these components approach the pressure-active and pressure-null parts of the present partition. The interference coefficient of figure~\ref{fig:sector_phase} measures this opposition directly, between the streaks generated by the two sectors of the leading RESPOD forcing mode.

	Destructive interference leaves the streak with less energy than the two sectors produce separately. With the mean values of $\chi_{\mathrm{s}}$ given above, the streak retains less than a quarter of the energy that the two sectors would deliver if they did not interfere. Writing $r=\|\hat{\boldsymbol q}_{0}\|/\|\hat{\boldsymbol q}_{\mathrm{a,s}}\|$ for the amplitude ratio and $\cos\varphi=\mathrm{Re}\langle \hat{\boldsymbol q}_{0},\hat{\boldsymbol q}_{\mathrm{a,s}}\rangle/(\|\hat{\boldsymbol q}_{0}\|\,\|\hat{\boldsymbol q}_{\mathrm{a,s}}\|)$ for the phase cosine, the definition (\ref{eq:chi_s}) becomes
	\begin{equation}
		\chi_{\mathrm{s}}=\frac{2r\cos\varphi}{1+r^{2}},
		\label{eq:chi_s_trig}
	\end{equation}
	so complete cancellation requires both exact antiphase and equal amplitudes. The phase cosine is negative in every wave. Averaged over Reynolds number, the two responses are about $35^\circ$ from antiphase at $AR=2$ and about $19^\circ$ at $AR=8$. This proximity to antiphase motivates representing the null response by a correction with a real rather than a complex coefficient. A real coefficient changes the amplitude of the response but not its phase, which is sufficient when the two responses are exactly opposed. We test in \S\ref{sec:model} whether such a single coefficient is enough.

	\subsection{Work done by the two sectors}
	\label{sec:source_sink}

	The interference coefficient compares the two sector responses with each other. The energy budget determines which sector supplies energy and which removes it, through the work that each sector of the forcing does on the response.

	The fluctuating kinetic energy at a given scale is supplied by production from the mean shear, redistributed among scales by the nonlinear term and removed by viscous dissipation. For one Fourier mode, the real part of the inner product of the linearised momentum equation with the response $\hat{\boldsymbol q}$ gives
	\begin{equation}
		-\mathrm{Re}\langle \hat u,U'\hat v\rangle
		+\mathrm{Re}\langle \hat{\boldsymbol q},\nu\nabla^{2}\hat{\boldsymbol q}\rangle
		+\mathrm{Re}\langle \hat{\boldsymbol q},\hat{\boldsymbol f}\rangle
		=0,
		\label{eq:energy_balance}
	\end{equation}
	where $U'=\mathrm{d}U/\mathrm{d}y$ is the mean shear, with a prime denoting differentiation with respect to $y$. The three terms are the production, the viscous term and the work of the nonlinear forcing. Pressure does no net work on the response when integrated over the channel height \citep{AubryEtAl1988}, so it does not appear in (\ref{eq:energy_balance}). The work of the forcing is positive at scales that receive energy from other scales and negative at scales that lose energy to them. Spectral budgets of this kind have been used to map the transfer between scales in channel flow \citep{Mizuno2016,ChoHwangChoi2018,LeeMoser2019,DingEtAl2025} and, in the resolvent context, to show that the nonlinear term acts as a net sink at the energetic scales \citep{SymonEtAl2021}. Applying the sector projectors splits this work into the contributions of the two sectors.

	\begin{figure}
		\centering
		\includegraphics[width=0.55\textwidth]{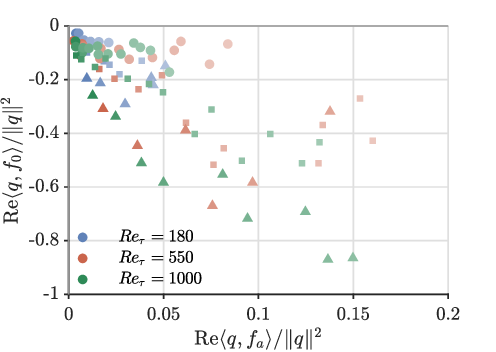}
		\caption{Work performed on the response by the pressure-null forcing against that performed by the pressure-active forcing, both per unit response energy and averaged over the realisations of each wave. One marker per wave: blue, $\Rey_\tau\approx180$; red, $550$; green, $1000$; circles $AR=2$, squares $AR=4$, triangles $AR=8$, with lighter shades for shorter $\lambda_x^+$. The lines through the origin divide the plane into quadrants; every wave lies in the fourth, $W_{\mathrm{a}}>0$ and $W_0<0$.}
		\label{fig:source_sink_blocks}
	\end{figure}

	We first evaluate this work on the full forcing data, without a modal decomposition. Figure~\ref{fig:source_sink_blocks} uses the measured response and forcing of each realisation and applies only the sector projectors. The forcing splits as the response does in (\ref{eq:sector_responses}),
	\begin{equation}
		\hat{\boldsymbol f}_{0}=\Bs\Pz\hat{\boldsymbol a},
		\qquad
		\hat{\boldsymbol f}_{\mathrm{a}}=\Bs\Pa\hat{\boldsymbol a},
		\qquad
		\hat{\boldsymbol a}=\Bs^{\dagger}\hat{\boldsymbol f},
		\label{eq:sector_forcing}
	\end{equation}
	and the work each part does on the response, per unit response energy, is
	\begin{equation}
		W_{\mathrm{a}}
		=
		\frac{\sum_{k}\mathrm{Re}\langle \hat{\boldsymbol q}^{(k)},\hat{\boldsymbol f}_{\mathrm{a}}^{(k)}\rangle}
		{\sum_{k}\|\hat{\boldsymbol q}^{(k)}\|^{2}},
		\qquad
		W_{0}
		=
		\frac{\sum_{k}\mathrm{Re}\langle \hat{\boldsymbol q}^{(k)},\hat{\boldsymbol f}_{0}^{(k)}\rangle}
		{\sum_{k}\|\hat{\boldsymbol q}^{(k)}\|^{2}},
		\label{eq:sector_work}
	\end{equation}
	where the sums run over the realisations of one wave and the projections (\ref{eq:sector_forcing}) are applied to each realisation. Being normalised by the response energy, $W_{\mathrm{a}}$ and $W_0$ have the dimensions of a growth rate and can be compared between waves of different energy. The parts $\hat{\boldsymbol f}_{0}$ and $\hat{\boldsymbol f}_{\mathrm{a}}$ add up to the solenoidal part of the measured forcing. The irrotational part produces no velocity and drops out of the projection, so $W_{0}+W_{\mathrm{a}}$ is the work of the measured forcing on the response. In the balance (\ref{eq:energy_balance}), positive work adds energy to the response and negative work removes it.

	In every wave, $W_{\mathrm{a}}>0$ and $W_0<0$. Averaged over the realisations, the active sector therefore supplies energy to the response and the null sector removes it at every Reynolds number, aspect ratio and wavelength. The same signs hold in most individual realisations. The active work is positive in at least $96$ per cent of the realisations at each Reynolds number, and the null work is negative in $83$ to $93$ per cent. The null sector removes more energy from the response than the active sector supplies to it. The median of $|W_0|/W_{\mathrm{a}}$ over the waves of each $(\Rey_\tau,AR)$ combination lies between $3.6$ and $7.6$, so the nonlinear term is a net sink at these wavenumbers. \citet{SymonEtAl2021} reported this net sink for the streamwise-constant scales. The sector decomposition separates it into a supply by the pressure-active forcing and a larger removal by the pressure-null forcing.

	These signs can be read through the balance (\ref{eq:energy_balance}), written separately for the streak $\hat{\boldsymbol q}_{\mathrm s}=\Smat\hat{\boldsymbol q}$ and the roll $\hat{\boldsymbol q}_{\mathrm r}=(\Imat-\Smat)\hat{\boldsymbol q}$ of the response. The active forcing has no streak component and the null forcing no roll component so that
	\begin{equation}
		-\frac{\beta}{k}\,\mathrm{Re}\langle \hat u_c,U'\hat v\rangle
		+\mathrm{Re}\langle \hat{\boldsymbol q}_{\mathrm s},\nu\nabla^{2}\hat{\boldsymbol q}_{\mathrm s}\rangle
		+\mathrm{Re}\langle \hat{\boldsymbol q},\hat{\boldsymbol f}_{0}\rangle
		=0
		\label{eq:streak_balance}
	\end{equation}
	for the streak and
	\begin{equation}
		-\frac{\alpha}{k}\,\mathrm{Re}\langle \hat u_k,U'\hat v\rangle
		+\mathrm{Re}\langle \hat{\boldsymbol q}_{\mathrm r},\nu\nabla^{2}\hat{\boldsymbol q}_{\mathrm r}\rangle
		+\mathrm{Re}\langle \hat{\boldsymbol q},\hat{\boldsymbol f}_{\mathrm a}\rangle
		=0
		\label{eq:roll_balance}
	\end{equation}
	for the roll. The first term of each equation is the production, which for the streak is the lift-up. The last terms are the sector works of (\ref{eq:sector_work}) before normalisation. A positive $W_{\mathrm{a}}$ therefore shows that the roll dissipates more energy than it receives from the mean shear, and is sustained by the nonlinear term. A negative $W_0$ shows that lift-up supplies more energy to the streak than viscosity dissipates, and the nonlinear term removes the difference. This is consistent with the self-sustaining process \citep{HamiltonEtAl1995,Waleffe1997}, in which the nonlinear term draws energy from the streak and regenerates the roll.

	\begin{figure}
		\centering
		\includegraphics[width=.8\textwidth]{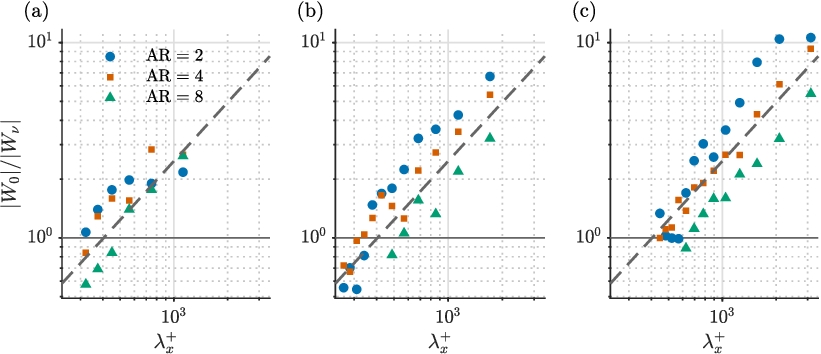}
		\caption{Magnitude of the null-sector work relative to the molecular viscous dissipation, $|W_0|/|W_\nu|$, against streamwise wavelength, at (a) $\Rey_\tau\approx180$, (b) $550$ and (c) $1000$. One marker per wave, taken as the median over realisations of the per-realisation ratio: blue circles $AR=2$, orange squares $AR=4$, green triangles $AR=8$. The horizontal line marks unity and the dashed line, identical in the three panels, has slope one. Both axes logarithmic.}
		\label{fig:null_work_viscous}
	\end{figure}

	To judge the size of this energy removal, we compare it with molecular dissipation in figure~\ref{fig:null_work_viscous}, which shows $|W_0|/|W_\nu|$. Here $W_\nu$ is the viscous term of (\ref{eq:energy_balance}) in the same normalisation as the sector works,
	\begin{equation}
		W_{\nu}
		=
		\frac{\sum_{k}\mathrm{Re}\langle \hat{\boldsymbol q}^{(k)},\nu\nabla^{2}\hat{\boldsymbol q}^{(k)}\rangle}
		{\sum_{k}\|\hat{\boldsymbol q}^{(k)}\|^{2}}.
		\label{eq:viscous_work}
	\end{equation}
	The ratio exceeds unity in most waves, with a median between $1.5$ and $2.1$ at each Reynolds number. The pressure-null forcing therefore removes energy at a rate comparable to, and generally larger than, molecular dissipation. The ratio also grows with streamwise wavelength, roughly in proportion to it. A power-law fit against $\lambda_x^+$ gives an exponent between $0.9$ and $1.3$, depending on whether the Reynolds numbers and aspect ratios are fitted together or separately. Including the streamwise fundamental, excluded in \S\ref{sec:database}, would lower this exponent but would leave the signs of $W_{\mathrm{a}}$ and $W_0$ unchanged. Both $|W_0|$ and $|W_\nu|$ decrease with wavelength, but molecular dissipation falls faster than the null-sector work. This suggests representing the null-sector removal by an added diffusion whose strength relative to molecular dissipation increases with the size of the structure. The model of \S\ref{sec:model} takes this form.

	\begin{figure}
		\centering
		\includegraphics[width=0.53\textwidth]{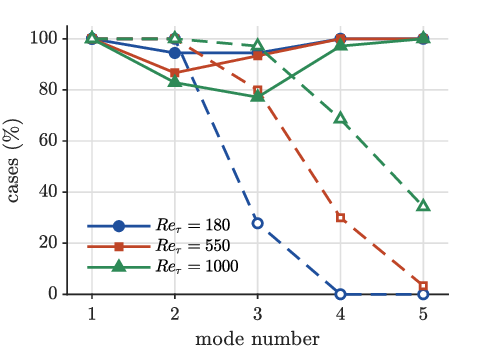}
		\caption{Percentage of waves satisfying each sign condition, against mode number. Solid lines with filled markers, the source condition $\mathrm{Re}\langle \hat{\boldsymbol q},\hat{\boldsymbol f}_{\mathrm{a}}\rangle>0$; dashed lines with open markers, the sink condition $\mathrm{Re}\langle \hat{\boldsymbol q},\hat{\boldsymbol f}_0\rangle<0$. Blue with circles, $\Rey_\tau\approx180$; red with squares, $550$; green with triangles, $1000$.}
		\label{fig:source_sink_modes}
	\end{figure}

	The source--sink analysis above uses the full solenoidal forcing of each realisation and the response it produces. We now examine how much of it carries over to the SPOD modes. Figure~\ref{fig:source_sink_modes} pairs each SPOD mode $\hat{\boldsymbol\psi}_j$ with the corresponding RESPOD mode $\hat{\boldsymbol a}_j$, which generates it through (\ref{eq:respod_closure}). In every wave, the leading mode satisfies both conditions. The active forcing supplies energy to it, and the null forcing removes energy from it. Beyond the leading mode, the source condition continues to hold in at least three quarters of the waves at every mode. The sink condition still holds in every wave at the second mode but becomes rarer beyond it, and it is lost earlier at $\Rey_\tau\approx180$ than at $1000$. The source role of the active sector thus extends to the subleading modes, whereas the sink role of the null sector is observed consistently only for the two leading modes. Since the subleading modes carry less energy and are estimated from the same number of realisations, part of the decay of the sink condition may reflect the poorer convergence of these modes rather than a change in the transfer itself.

	\subsection{Dependence of the opposition on streak amplitude}
	\label{sec:saturation}

	The results above are obtained from Fourier transforms over blocks of $N_{\mathrm{fft}}$ samples, averaged under the assumption of statistical stationarity, and can therefore hide a dependence on the instantaneous amplitude. To test for it, we condition the time histories $\boldsymbol q(t)$ and $\boldsymbol f(t)$ of each wave on the instantaneous streak energy, without a temporal Fourier transform. We obtain the streak $\boldsymbol q_{\mathrm s}(t)=\Smat\boldsymbol q(t)$ and the null forcing $\boldsymbol f_0(t)$ with the projections (\ref{eq:streak_projector}) and (\ref{eq:projectors}), which depend on $(\alpha,\beta)$ only and are applied at each instant. The alignment between the null forcing and the streak,
	\begin{equation}
		A=-\frac{\mathrm{Re}\langle \boldsymbol q_{\mathrm{s}},\boldsymbol f_0\rangle}
		{\|\boldsymbol q_{\mathrm{s}}\|\,\|\boldsymbol f_0\|},
		\qquad
		D_{\rm sat}
		=-\frac{\mathrm{Re}\langle \boldsymbol q_{\mathrm{s}},\boldsymbol f_0\rangle}
		{\|\boldsymbol q_{\mathrm{s}}\|^{2}}
		=A\,\frac{\|\boldsymbol f_0\|}{\|\boldsymbol q_{\mathrm{s}}\|},
		\label{eq:saturation_metrics}
	\end{equation}
	is positive when the forcing opposes the streak, and $D_{\rm sat}$ is the corresponding work removed per unit streak energy. Writing $D_{\rm sat}$ as the product of $A$ and the amplitude ratio separates the two ways it can change.

	\begin{figure}
		\centering
		\includegraphics[width=.85\textwidth]{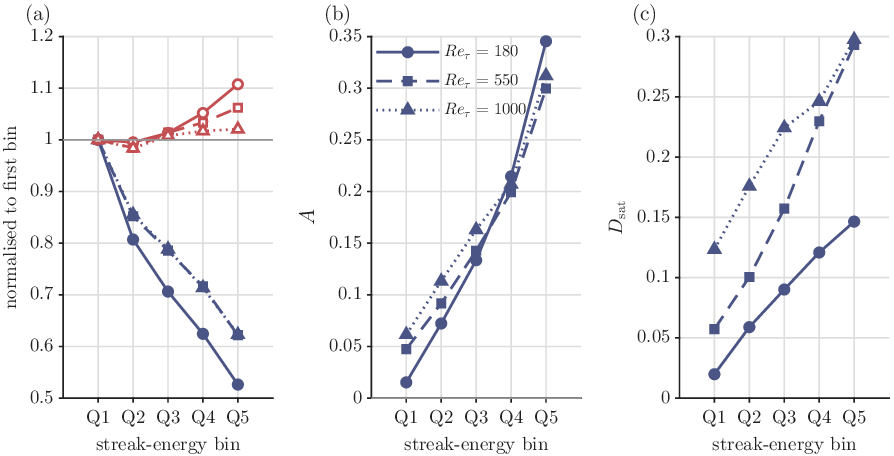}
		\caption{Quantities conditioned on the instantaneous streak energy, in five bins of increasing energy. (a) Filled markers and dark blue, $\|\boldsymbol f_0\|/\|\boldsymbol q_{\mathrm{s}}\|$; open markers and red, $\|\boldsymbol f_0\|$; both normalised by their value in the first bin, with the horizontal line at unity. (b) Alignment $A$ between the null forcing and the streak. (c) Work removed per unit streak energy, $D_{\rm sat}$. In all panels solid lines with circles denote $\Rey_\tau\approx180$, dashed lines with squares $550$, and dotted lines with triangles $1000$.}
		\label{fig:time_saturation}
	\end{figure}

	Figure~\ref{fig:time_saturation} sorts the time samples of each wave into five bins of increasing streak energy, with the same number of samples in each. Each curve shows the median, over the waves at one Reynolds number, of the bin medians. Across those bins the streak energy rises by a factor of $4.4$ at $\Rey_\tau\approx180$ and $2.7$ at $1000$, while the null forcing magnitude $\|\boldsymbol f_0\|$ changes by at most $11$ per cent. The magnitude of the null forcing is therefore nearly independent of the streak amplitude. The ratio $\|\boldsymbol f_0\|/\|\boldsymbol q_{\mathrm{s}}\|$ falls accordingly, because its denominator grows while its numerator stays nearly constant.

	Despite this falling amplitude ratio, $D_{\rm sat}$ increases across the bins, as figure~\ref{fig:time_saturation}(c) shows. Since $D_{\rm sat}=A\,\|\boldsymbol f_0\|/\|\boldsymbol q_{\mathrm{s}}\|$, the increase must come from the alignment. As figure~\ref{fig:time_saturation}(b) shows, $A$ rises from nearly zero in the lowest bin to about $0.3$ in the highest, which is a larger relative change than the decrease in the amplitude ratio. The same trend is found without binning. We computed the Spearman rank correlation, which measures whether one quantity increases monotonically with the other, over the time samples of each wave. It is positive in every wave, with a median of $0.31$ between streak energy and $A$ and $0.18$ between streak energy and $D_{\rm sat}$. Because $A$ and $D_{\rm sat}$ are evaluated at the same instant as the streak energy, without a time lag, the conditioning shows that a higher streak energy and a closer alignment occur together, without indicating which follows the other.

	\begin{figure}
		\centering
		\includegraphics[width=.8\textwidth]{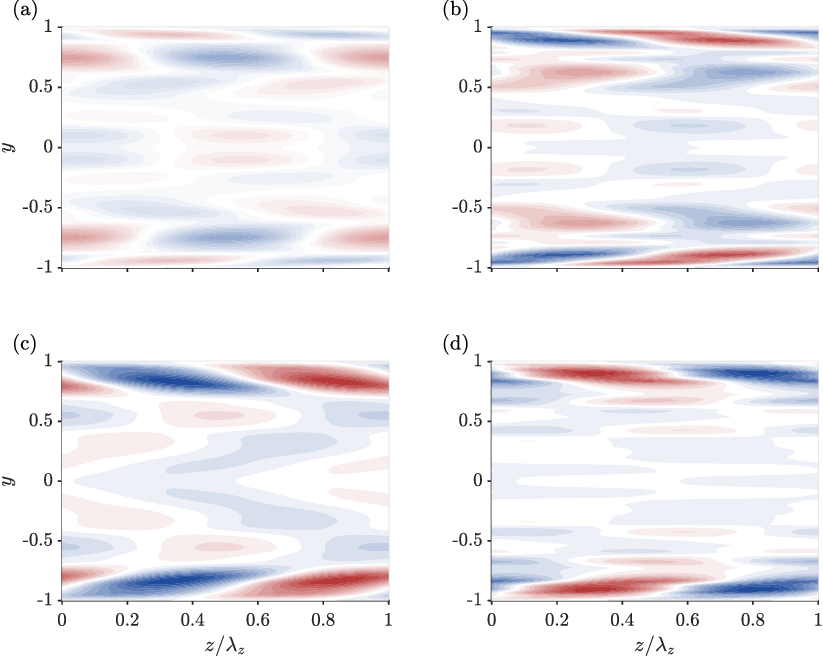}
		\caption{Instantaneous fields in the cross-stream plane at two instants, $\Rey_\tau\approx180$, $AR=4$, $\lambda_x^+=321$. Top row (a,b), an instant at the tenth percentile of streak energy; bottom row (c,d), the ninetieth. Left column (a,c), the streak response $\boldsymbol q_{\mathrm{s}}$; right column (b,d), the pressure-null forcing $\boldsymbol f_0$. Both fields lie along the crests and are shown through their crest component, red and blue denoting the two signs. The colour scale is symmetric about zero, shared down each column and differing between columns.}
		\label{fig:streak_snapshots}
	\end{figure}

	The increase in alignment can be seen in the instantaneous fields of a single wave, shown in figure~\ref{fig:streak_snapshots} at a low-energy and a high-energy instant. We chose this wave because the change in $\|\boldsymbol f_0\|$ between its two instants is the closest to the median over all waves at $\Rey_\tau\approx180$. It is seen that the forcing magnitude is similar in the two rows, with $\|\boldsymbol f_0\|$ changing by a factor of $1.07$, whereas $\|\boldsymbol q_{\mathrm{s}}\|$ rises by a factor of $2.06$. At the high-energy instant the null forcing opposes the streak across the channel, whereas at the low-energy instant the two are less systematically related.

	As the streak energy increases, the pressure-null forcing therefore keeps a nearly constant amplitude but aligns more closely against the streak, making the cancellation more efficient. Opposition of the streak by the nonlinear term is one stage of the self-sustaining process of \citet{HamiltonEtAl1995} and \citet{Waleffe1997}, first characterised in minimal flow units that sustain a single structure. Controlled numerical experiments have shown that the same process operates outside such units. \citet{HwangCossu2010c} isolated it at large scales by damping the smaller ones, and \citet{deGiovanettiEtAl2017} identified the streak instability that seeds large-scale motion by driving a streak with a body force computed from linear theory. Both studies intervene in the flow to isolate the mechanism. We conduct the present measurement in unforced turbulence, with the nonlinear forcing taken as it stands from the simulation and separated by the sector projectors. We find that the opposition holds in every wavenumber pair, at three Reynolds numbers and over a decade in streamwise wavelength.

	One should note that the alignment $A$ does not measure the same opposition as the phase cosine of (\ref{eq:chi_s_trig}). The phase cosine compares the response to the null forcing, $\boldsymbol q_0$, with the active streak, whereas $A$ compares the null forcing with the total streak $\boldsymbol q_{\mathrm s}=\boldsymbol q_0+\boldsymbol q_{\mathrm{a,s}}$. Splitting the streak in the numerator of $A$ gives
	\begin{equation}
		-\mathrm{Re}\langle\boldsymbol q_{\mathrm s},\boldsymbol f_0\rangle
		=-\mathrm{Re}\langle\boldsymbol q_{\mathrm{a,s}},\boldsymbol f_0\rangle
		-\mathrm{Re}\langle\boldsymbol q_0,\boldsymbol f_0\rangle .
	\end{equation}
	The first term on the right hand side is the opposition of the null forcing to the active streak. The second is the work of the null forcing on its own response. This work is positive on average, because the null forcing is the only source of energy of $\boldsymbol q_0$ and has to supply what viscosity dissipates. With its minus sign, the second term is therefore negative and reduces $A$. We do not evaluate the two terms separately, because $\boldsymbol q_0$ is not available at each instant. The resolvent gives $\boldsymbol q_0$ only from the Fourier transform of a block of $N_{\mathrm{fft}}$ samples, which removes the instantaneous resolution that the conditioning requires. Obtaining $\boldsymbol q_0(t)$ directly would require integrating the linearised equations from an unknown initial state, whose influence could outlast the stored time histories.

	The measurements of \S\ref{sec:source_sink_group} together show that the pressure-null forcing is correlated with the response in a consistent way. It opposes the response in phase and removes energy from it at a rate comparable to molecular dissipation, and this rate relative to molecular dissipation grows with the size of the structure. These are statistical relations and do not show that the forcing is produced by the response. They are, however, the properties that a diffusion acting on the response would have. The next section uses this observation to model the pressure-null forcing.

	\section{An eddy-viscosity model of the pressure-null forcing}
	\label{sec:model}

	As discussed in \S\ref{sec:introduction}, part of the effect of the unresolved turbulence on a coherent structure can be represented by an eddy viscosity added to the linear operator, usually with the profile of \citet{Cess1958}, which was designed for the mean Reynolds shear stress. Here, we adopt this strategy to model the pressure-null sector of the forcing. We first choose the profile and the amplitude of the eddy viscosity by comparing the null part of the force it implies with the measured pressure-null forcing (\S\ref{sec:model_selfsimilar} and \S\ref{sec:model_profiles}). We then test this eddy viscosity in the conventional operator, where it acts on the whole response (\S\ref{sec:model_autonomous}). Finally, we keep the same profile and amplitude, confine the eddy viscosity to the pressure-null sector, and compare the two operators when both are driven by the measured pressure-active forcing (\S\ref{sec:model_closed_loop}).

	An eddy viscosity enters the resolvent through an increment $\delta\nu(y)$ added to the molecular viscosity in all momentum equations, so that $\nu$ is replaced by $\nu+c\,\delta\nu$ in the operator that generates $\Hu$ in (\ref{eq:primitive_resolvent}), with $c$ a real amplitude. The added diffusion is
	\begin{equation}
		\mathrm{d}\mathsfbi V=\mathrm{diag}(\delta\nu)\left(D^{2}-k^{2}\right)
		+\mathrm{diag}(\delta\nu')D,
		\label{eq:dV}
	\end{equation}
	and the standard choice is $\delta\nu=\nu_{t}-\nu$, with $\nu_{t}(y)$ the eddy-viscosity profile of \citet{Cess1958}, as given by \citet{ReynoldsTiederman1967},
	\begin{equation}
		\frac{\nu_t}{\nu}=\frac12\left\{1+\frac{\kappa^2\Rey_\tau^2}{9}\,(1-y^2)^2(1+2y^2)^2
		\Big[1-\mathrm{e}^{(|y|-1)\Rey_\tau/A_{\mathrm{d}}}\Big]^2\right\}^{1/2}+\frac12,
		\label{eq:cess}
	\end{equation}
	with $y$ in units of $H$, $\Rey_\tau$ obtained from the wall gradient of the mean profile, and $\kappa=0.426$ and $A_{\mathrm{d}}=25.4$, which \citet{delAlamoJimenez2006} fitted to DNS at $\Rey_\tau\approx2000$. The added term is equivalent to a force that depends linearly on the velocity, $c\,\mathrm{d}\mathsfbi V\hat{\boldsymbol q}$, fed back into the molecular resolvent. Writing that force in the solenoidal coordinates on which $\Rs$ acts gives the eddy-viscosity resolvent
	\begin{equation}
		\Rev(c)=\big(\Imat-c\,\Rs\Bs^{\dagger}\,\mathrm{d}\mathsfbi V\big)^{-1}\Rs.
		\label{eq:Rev}
	\end{equation}
	Here $\Bs^{\dagger}$ of (\ref{eq:Bs_dagger}) maps the force field onto the solenoidal forcing coordinates, and $\mathrm{d}\mathsfbi V$ acts on the velocity field of the response.

	We use the null part of this force,
	\begin{equation}
		\hat{\boldsymbol f}_{\nu}=c\,\Bs\Pz\Bs^{\dagger}\,\mathrm{d}\mathsfbi V\hat{\boldsymbol q},
		\label{eq:ev_null_forcing}
	\end{equation}
	as the model of the pressure-null forcing. Evaluated on the leading SPOD mode, $\hat{\boldsymbol q}=\hat{\boldsymbol\psi}_1$, it should approximate the measured null forcing $\hat{\boldsymbol f}_{0}=\Bs\Pz\hat{\boldsymbol a}_1$ of (\ref{eq:sector_forcing}) as closely as possible. We seek the approximation for the leading mode, not in each realisation, so it is a statement about second-order statistics. We measure shape agreement between two vectors throughout by the weighted modal assurance criterion, formed with the inner product of (\ref{eq:L2_norm}),
\begin{equation}
\operatorname{MAC}(\boldsymbol{x},\boldsymbol{y})
=
\frac{|\langle\boldsymbol{x},\boldsymbol{y}\rangle|^2}{\|\boldsymbol{x}\|^2\,\|\boldsymbol{y}\|^2},
\label{eq:weighted_mac}
\end{equation}
	which is invariant to the amplitude and phase of either vector.

	\subsection{Wall-normal profiles of the pressure-null forcing}
	\label{sec:model_selfsimilar}

	The amplitude and the wall-normal extent of the pressure-null forcing of the leading mode vary from wave to wave, mainly with its wavelength in wall units and its aspect ratio. To simplify the comparison, we scale the wall distance with the centre of the forcing, following the self-similar coordinate of \citet{KarbanEtAl2022}. Because $\hat{\boldsymbol f}_{0}$ has no wall-normal component, its magnitude is $|\hat{\boldsymbol f}_{0}|=\big(|\hat f_{0,x}|^{2}+|\hat f_{0,z}|^{2}\big)^{1/2}$. The centre $y_h$ is the half-area point of this magnitude over the half channel,
	\begin{equation}
		\int_{0}^{y_h}|\hat{\boldsymbol f}_{0}|\,\mathrm{d}y=\frac{1}{2}\int_{0}^{H}|\hat{\boldsymbol f}_{0}|\,\mathrm{d}y,
		\label{eq:yh}
	\end{equation}
	with $y$ measured from the wall, and we compare the profiles as functions of $\eta=y/y_h$ after normalising each by its peak.

\begin{figure}
		\centering
		\includegraphics[width=.85\textwidth]{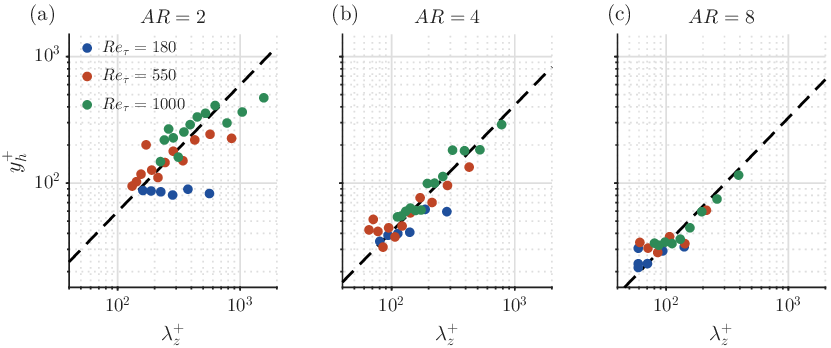}
		\caption{Centre $y_h^+$ of the pressure-null forcing of the leading mode, from (\ref{eq:yh}) in wall units, against $\lambda_z^+$ for (a) $AR=2$, (b) $4$ and (c) $8$, coloured by Reynolds number. Dashed line, $y_h\propto\lambda_z$ through the median of $y_h/\lambda_z$ in each panel.}
		\label{fig:yh_lambdaz}
	\end{figure}

	The centre follows the attached-eddy scaling $y_h\propto\lambda_z$ except for the largest waves at $AR=2$ and the structures closest to the wall (figure~\ref{fig:yh_lambdaz}). The height of the structures, measured by $y_h$, also depends on the aspect ratio. For the same $\lambda_z$ and Reynolds number, the forcing of the structure with the lower aspect ratio lies farther from the wall. The half-area point measures the height of a mode that does not reach the channel centre. Once the mode extends to the centre, $y_h$ reaches a plateau below $0.5H$. This is seen to affect the $AR=2$ waves with the largest wavelengths in outer units. At the other end of the size range, for the smallest waves at $AR=4$ and $8$, with $\lambda_z^+<100$, the scaling would place the centre within the buffer layer. Their null forcing peaks at $y^+$ between $14$ and $22$ at all Reynolds numbers, so their height is set by the wall in viscous units and not by $\lambda_z$ \citep{KarbanEtAl2022}. Note that the self-similar coordinate is used here only to display the profiles of different waves together. We do not build the null forcing model on it.

	\begin{figure}
		\centering
		\includegraphics[width=.85\textwidth]{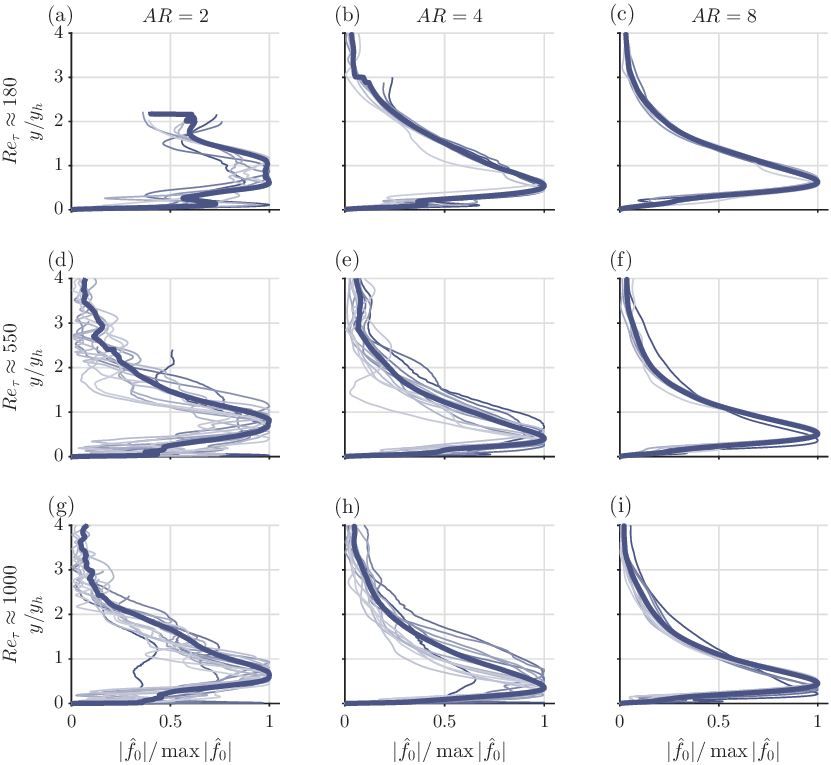}
		\caption{Wall-normal profiles of the pressure-null forcing of the leading mode, $|\hat{\boldsymbol f}_0|$ normalised by its peak, against $\eta=y/y_h$, at (a--c) $\Rey_\tau\approx180$, (d--f) $550$ and (g--i) $1000$, for $AR=2$, $4$ and $8$ from left to right. Thin lines, individual waves, lighter for shorter $\lambda_x^+$; thick line, mean over the waves of the panel.}
		\label{fig:null_selfsimilar}
	\end{figure}

	Figure~\ref{fig:null_selfsimilar} shows the profiles of all waves in the scaled coordinate. Each curve ends at the channel centre, which lies at $\eta=H/y_h$ and so at a different $\eta$ for each wave. Where some curves reach the centre before decaying to zero, we observe sudden changes in the mean over the waves. This is seen in the $AR=2$ waves at $\Rey_\tau\approx180$, whose curves end at $\eta\approx2$, and for the largest $AR=4$ waves near $\eta\approx3$. These changes result from the averaging and not from the forcing, and should be disregarded. The profiles within a panel differ less as the aspect ratio increases. At $AR=8$, individual waves follow the mean profile closely, while much larger differences are seen at $AR=2$. Apart from $AR=2$ at $\Rey_\tau\approx180$, the mean profiles peak at a similar height, around $\eta\approx0.5$, and decay similarly at every aspect ratio and Reynolds number. We therefore compare the eddy-viscosity profiles of \S\ref{sec:model_profiles} with these mean profiles, so that one comparison covers waves of different size, while we measure the agreement of each wave separately by the MAC.

	\subsection{Profile and amplitude of the eddy viscosity}
	\label{sec:model_profiles}

\begin{figure}
		\centering
		\includegraphics[width=.85\textwidth]{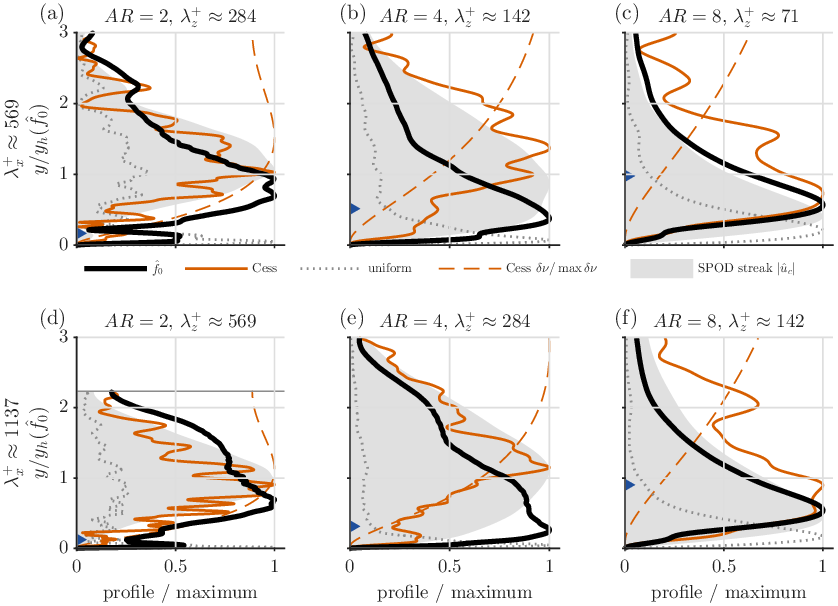}
		\caption{Null forcing of single waves at $\Rey_\tau\approx550$ against $\eta=y/y_h(\hat{\boldsymbol f}_0)$, for (a--c) $\lambda_x^+\approx570$ and (d--f) $\lambda_x^+\approx1140$, with $AR=2$, $4$ and $8$ from left to right. Black, measured null forcing $\hat{\boldsymbol f}_0$; orange solid, the null forcing (\ref{eq:ev_null_forcing}) implied on the measured leading mode by the Cess profile; grey dotted, by a uniform eddy viscosity; orange dashed, the Cess profile, shown as $\nu_t-\nu$; shaded, the streak $|\hat u_c|$ of the leading SPOD mode. All curves are normalised by their maxima. The horizontal grey line in panel (d) marks the channel centre. The triangle on the vertical axis marks $y^+=30$, the upper edge of the buffer layer.}
		\label{fig:null_forcing_waves}
	\end{figure}

	We evaluate the null forcing implied by an eddy-viscosity profile with (\ref{eq:ev_null_forcing}) on the measured leading mode, $\hat{\boldsymbol q}=\hat{\boldsymbol\psi}_1$, and compare it with $\hat{\boldsymbol f}_{0}$ in the coordinate $\eta$ of the measured forcing. This tests the profile alone, since the amplitude $c$ drops out of the normalised profile and of the MAC. Figure~\ref{fig:null_forcing_waves} shows single waves at $\Rey_\tau\approx550$, two streamwise wavelengths for each aspect ratio. It compares the measured null forcing with the null forcing implied by the Cess profile and by a uniform eddy viscosity, together with the Cess profile itself and the streak of the leading SPOD mode. The modelled forcing contains second derivatives of the SPOD mode, which amplify its statistical noise and cause the wiggles in the modelled profiles.

	Figure~\ref{fig:null_forcing_waves} suggests that the null forcing peaks below the streak core when the core lies above the buffer layer. In panels (a,b,d,e) the streak is centred above $y^+=30$, the upper edge of this layer, which is marked on the vertical axis, and the forcing peaks below its core. In panels (c,f), where the core lies in the buffer layer, the forcing is centred at the core. The null forcing of a tall streak is therefore concentrated in its lower part, and that of a streak in the buffer layer at its core. This is consistent with \citet{ChoHwangChoi2018}, who found that the energy at a spanwise scale $\lambda_z$ is produced, and transferred to smaller scales by the nonlinear terms, mainly at $y\approx\lambda_z/5$. At $AR=4$ and $8$ the measured forcing peaks close to this height, and at $AR=2$ about twice as high.

	A uniform eddy viscosity, $\delta\nu$ constant, gives a null forcing proportional to $\Delta\hat u_c$, the Laplacian of the streak. The streak falls to zero over a thin layer at the wall, which makes its curvature $D^2\hat u_c$, and hence $\Delta\hat u_c$, largest there. The Cess profile removes this near-wall contribution through the damping of \citet{vanDriest1956} that it contains, but above the buffer layer it keeps growing as the mixing length $\kappa y$, through the core of the streak and beyond. The null forcing given by the Cess profile is then weighted towards the outer part of the streak, which it overdamps, and it peaks above the measured forcing in every panel of figure~\ref{fig:null_forcing_waves}. At $AR=2$ the structures extend farther from the wall (figure~\ref{fig:yh_lambdaz}), and the Cess profile remains nearly uniform above the height at which the forcing it implies reaches its peak. At this AR, the Cess forcing is therefore similar to that of a uniform eddy viscosity, except the near-wall peak of the latter (figure~\ref{fig:null_forcing_waves}a,d).

	\citet{GuptaEtAl2021} corrected this outer excess by multiplying the eddy viscosity by $\lambda/(\lambda+\lambda_m)$, with $\lambda=2\pi/k$ the wavelength of the structure and $\lambda_m$ a lower bound for the size of the energy-containing eddies at each height, which grows linearly with $y$ in the logarithmic region. For a given structure this factor decreases with height and falls below one half where $\lambda_m$ exceeds the wavelength of the structure. \citet{YingEtAl2024} instead optimised the eddy-viscosity profile so that the resolvent modes align with the SPOD modes. The optimised profiles follow Cess near the wall and level off at a value that depends mostly on the spanwise wavelength. From these profiles they built a model that caps the Cess profile at a fraction of its maximum, $\delta\nu=\min\{\nu_t-\nu,\ \check\nu_m\max_y(\nu_t-\nu)\}$, with $\check\nu_m$ a fitted function of $\lambda_z/H$, and smooths the corner with a filter.

	Here, we adopt the construction of \citet{YingEtAl2024} and cap the Cess profile above a height proportional to the spanwise wavelength $\lambda_z$, as the centre of the measured forcing is (figure~\ref{fig:yh_lambdaz}),
	\begin{equation}
		\delta\nu(y)=\mathcal S\Big[\min\big\{\nu_t(y)-\nu,\ \nu_t(y_c)-\nu\big\}\Big],
		\qquad y_c=\max\{\theta\lambda_z,\ y_{\min}\}.
\label{eq:capped_cess}
\end{equation}
Here $\theta$ and $y_{\min}$ are fixed below, and $\mathcal S$ is the filter of \citet{YingEtAl2024},
	\begin{equation}
		\mathcal S[g]=\arg\min_{s}\int_{-H}^{H}\Big[(s-g)^2+\epsilon\,\tilde y^{\,4}\,(D^{2}s)^{2}\Big]\,\mathrm dy,
		\qquad \epsilon=5\times10^{-3},
		\label{eq:smoothing}
	\end{equation}
	with $\tilde y$ the distance to the nearest wall in units of $H$. The weight $\tilde y^4$ leaves the near-wall growth of Cess unchanged and rounds the corner over about a quarter of the cap height.

	The cap sets the height at which the modelled forcing peaks. Written along $\boldsymbol e_c$, the forcing (\ref{eq:ev_null_forcing}) is $c\,[\delta\nu\,\Delta\hat u_c+\delta\nu'D\hat u_c]$. The first term is the diffusion of the streak, and the second arises because $\delta\nu$ varies with $y$. Below the cap $\delta\nu$ is small, and under the core of the streak the second term has the opposite sign to the first and reduces the forcing. Above the cap $\delta\nu$ is constant, the second term vanishes and the forcing follows the diffusion of the streak. The forcing therefore peaks between the cap and the core of the streak, and a lower cap moves the peak towards the wall. We place the cap at the height where a structure loses most of its energy to smaller scales, which is $y\approx\lambda_z/5$ according to \citet{ChoHwangChoi2018} and agrees with the measured forcing at $AR=4$ and $8$ (figure~\ref{fig:null_forcing_waves}). This gives $\theta=0.2$. For the smallest waves the measured forcing peaks at a nearly fixed height in wall units (\S\ref{sec:model_selfsimilar}), and we use $y_{\min}^+=15$ as the lower bound of the cap height. We use the same values for all combinations.

	\begin{figure}
		\centering
		\includegraphics[width=.85\textwidth]{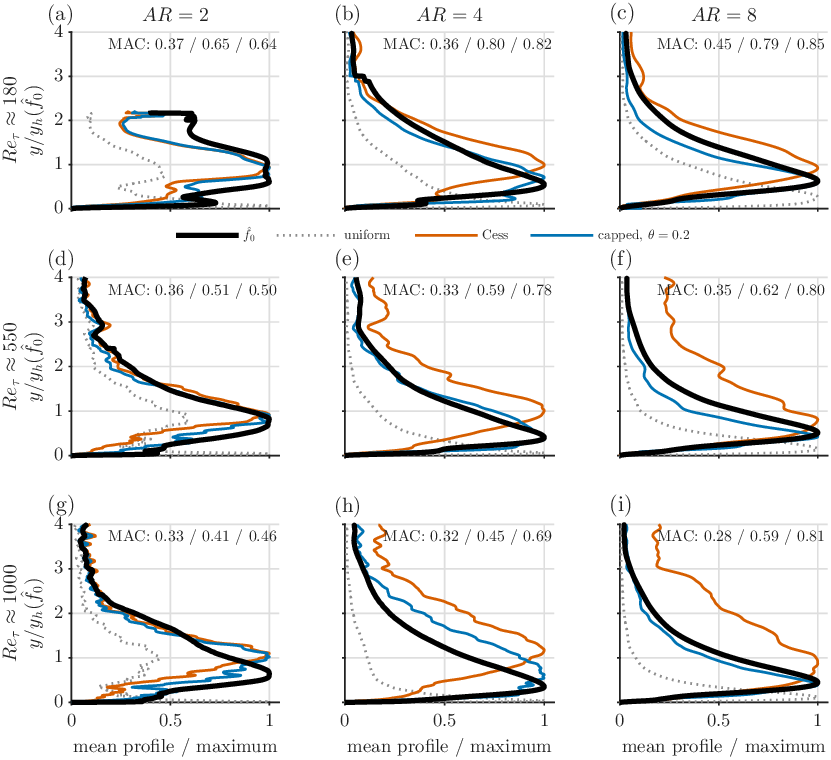}
		\caption{Mean profiles of the null forcing against $\eta=y/y_h(\hat{\boldsymbol f}_0)$, panels as in figure~\ref{fig:null_selfsimilar}. The measured null forcing $\hat{\boldsymbol f}_0$ is compared with the null forcing (\ref{eq:ev_null_forcing}) implied on the measured leading mode by a uniform eddy viscosity, by the Cess profile and by the capped profile (\ref{eq:capped_cess}) with $\theta=0.2$. The three values in each panel are the median MAC with $\hat{\boldsymbol f}_0$ for the uniform, Cess and capped profiles, in this order.}
		\label{fig:null_profiles}
	\end{figure}

	\begin{table}
		\centering
		\begin{tabular}{llcccccc}
			\toprule
			& & \multicolumn{3}{c}{MAC with $\hat{\boldsymbol f}_0$} & \multicolumn{3}{c}{$y_h/y_h(\hat{\boldsymbol f}_0)$}\\
			\cmidrule(lr){3-5}\cmidrule(lr){6-8}
			$\Rey_\tau$ & $AR$ & uniform & Cess & capped & uniform & Cess & capped\\
			\midrule
				$180$ & $2$ & $0.37$ & $0.65$ & $0.64$ & $0.72$ & $0.99$ & $0.95$\\
				 & $4$ & $0.36$ & $0.80$ & $0.82$ & $0.52$ & $1.15$ & $0.91$\\
				 & $8$ & $0.45$ & $0.79$ & $0.85$ & $0.51$ & $1.22$ & $0.81$\\
				\midrule
				$550$ & $2$ & $0.36$ & $0.51$ & $0.50$ & $0.95$ & $1.20$ & $1.06$\\
				 & $4$ & $0.33$ & $0.59$ & $0.78$ & $0.42$ & $1.64$ & $0.84$\\
				 & $8$ & $0.35$ & $0.62$ & $0.80$ & $0.39$ & $2.74$ & $0.82$\\
				\midrule
				$1000$ & $2$ & $0.33$ & $0.41$ & $0.46$ & $0.98$ & $1.15$ & $1.06$\\
				 & $4$ & $0.32$ & $0.45$ & $0.69$ & $0.55$ & $1.90$ & $1.09$\\
				 & $8$ & $0.28$ & $0.59$ & $0.81$ & $0.34$ & $5.19$ & $0.88$\\
			\bottomrule
		\end{tabular}
		\caption{Null forcing implied by three eddy-viscosity profiles, evaluated on the measured leading mode, against the measured null forcing. Median MAC (\ref{eq:weighted_mac}) and median ratio of the centre $y_h$ to that of the measured forcing over the waves of each combination. Capped, (\ref{eq:capped_cess}) with $\theta=0.2$.}
		\label{tab:profiles}
	\end{table}

	Figure~\ref{fig:null_profiles} compares the null forcing implied by the three eddy-viscosity profiles with the measured forcing, as mean profiles over the waves of each panel in the coordinate $\eta$. Table~\ref{tab:profiles} gives the corresponding medians of the MAC and of the forcing centre, which are evaluated for each wave separately. The uniform forcing peaks close to the wall in every panel, and the Cess forcing about twice as far from the wall as the measured forcing. The Cess forcing also extends farther from the wall. The ratio of its centre $y_h$ to that of the measured forcing (table~\ref{tab:profiles}) grows with aspect ratio and, at $AR=4$ and $8$, with Reynolds number, and reaches five at $\Rey_\tau\approx1000$ and $AR=8$. Both trends have the same origin. The Cess eddy viscosity increases with distance from the wall up to about half-way to the channel centre, whatever the size of the structure, while the measured forcing stays at a height set by $\lambda_z$. The smaller the structure is compared with the channel half-height, the more the eddy viscosity increases across and above it, and the more the modelled forcing is weighted towards the outer part of the streak. The capped forcing lies at the height of the measured forcing in every panel. At $AR=2$, it nearly coincides with the Cess forcing. The MAC is lowest for the uniform profile and highest for the capped profile, whose pooled median is $0.72$ against $0.58$ for Cess. The cap improves on Cess in nearly every wave at $\Rey_\tau\approx550$ and $1000$. At $\Rey_\tau\approx180$ the Cess forcing already lies close to the measured forcing, and the two profiles give similar MACs. Note that the MAC values at $AR=2$ are particularly low for the Cess and capped profiles, despite the agreement seen in the mean profiles of figure~\ref{fig:null_profiles}a,d,g. The MAC is evaluated on the complex forcing of each wave. An additional analysis, not shown here, indicates that at $AR=2$ most of the difference lies in the phase of the forcing across the layer and in small-scale oscillations of the individual profiles, and not in their overall shape. The leading SPOD mode is also least separated from the second at $AR=2$, so the convergence of the modes may contribute to lower MAC.

	These results depend only weakly on $\theta$. With $\theta=0.3$ and the other model constants unchanged, the mean MAC of the predicted leading mode (\S\ref{sec:model_autonomous}) decreases from $0.90$ to $0.88$, with the largest decrease, $0.04$, at $AR=8$. The value of $\theta$ that maximises the MAC with $\hat{\boldsymbol f}_{0}$ wave by wave has a median between $0.2$ and $0.3$ in eight of the nine combinations.

\begin{figure}
		\centering
		\includegraphics[width=.66\textwidth]{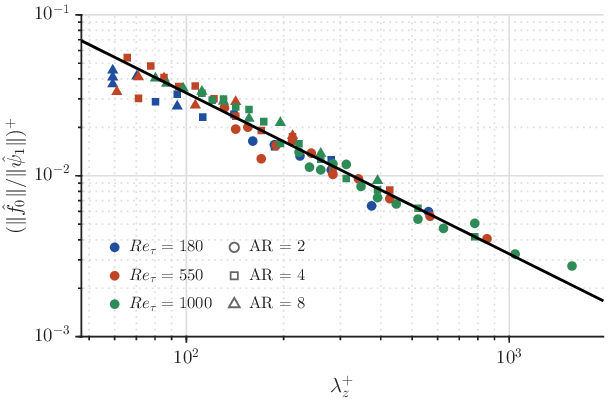}
		\caption{Amplitude of the null forcing per unit response, $(\|\hat{\boldsymbol f}_0\|/\|\hat{\boldsymbol\psi}_1\|)^+$, against $\lambda_z^+$, one marker per wave. Colours denote $\Rey_\tau$ and markers $AR$ as in figure~\ref{fig:phase_speed}. The line is (\ref{eq:amplitude_law}). Both axes logarithmic.}
		\label{fig:amplitude_law}
	\end{figure}

	The capped profile (\ref{eq:capped_cess}) fixes the shape of the modelled null forcing (\ref{eq:ev_null_forcing}), and the coefficient $c$ sets its amplitude. We now determine $c$ from the amplitude of the measured null forcing per unit response, $\|\hat{\boldsymbol f}_{0}\|/\|\hat{\boldsymbol\psi}_1\|$. Figure~\ref{fig:amplitude_law} shows that this amplitude decreases almost inversely with the spanwise wavelength, regardless of the Reynolds number and the aspect ratio. A fit over all waves gives
	\begin{equation}
		\left(\frac{\|\hat{\boldsymbol f}_{0}\|}{\|\hat{\boldsymbol\psi}_1\|}\right)^{+}
		=\frac{\gamma}{\lambda_z^{+}},
		\qquad \gamma=3.3.
		\label{eq:amplitude_law}
	\end{equation}

	With such a global trend, we fix $c$ by imposing the same relation on the model. In (\ref{eq:amplitude_law}) we replace $\hat{\boldsymbol\psi}_1$ by the leading response mode $\hat{\boldsymbol q}(c)$ of $\Rev(c)$, and $\hat{\boldsymbol f}_0$ by the null forcing (\ref{eq:ev_null_forcing}) that the model implies for this mode. The coefficient is then the root of
	\begin{equation}
		\left(c\,\frac{\|\Pz\Bs^{\dagger}\,\mathrm{d}\mathsfbi V\hat{\boldsymbol q}(c)\|}{\|\hat{\boldsymbol q}(c)\|}\right)^{+}
		=\frac{\gamma}{\lambda_z^{+}}.
		\label{eq:selfconsistent_c}
	\end{equation}
	Note that the only input in this relation from the DNS is the constant $\gamma$ and the mean flow. The left-hand side is zero at $c=0$ and increases monotonically with $c$ in every wave, so (\ref{eq:selfconsistent_c}) has a single root, which can be found numerically. Over all waves, $c$ lies between $0.46$ and $2.78$. It increases with aspect ratio and decreases with $\lambda_z$ at fixed aspect ratio, without a clear trend in $\Rey_\tau$.

	\subsection{Autonomous prediction of the leading mode}
	\label{sec:model_autonomous}

	With (\ref{eq:Rev}), (\ref{eq:capped_cess}) and (\ref{eq:selfconsistent_c}), we construct the eddy-viscosity resolvent from the mean profile alone, with no further input from the DNS. It is evaluated at the selected frequency $\omega_{\max}$ of each wave. The response is predicted by the singular value decomposition of $\Rev(c)$. We assess the prediction by the MAC between the leading left singular vector and the leading SPOD mode $\hat{\boldsymbol\psi}_1$, and by how close the gain separation $\sigma_1/\sigma_2$, the ratio of the two leading singular values of $\Rev(c)$, is to the SPOD value $\sqrt{\lambda_1/\lambda_2}$. We compare the capped profile, with $c$ from (\ref{eq:selfconsistent_c}), against the Cess profile at $c=1$, which is the standard choice. To isolate the effect of (\ref{eq:selfconsistent_c}), we also use the capped profile at $c=1$.

	\begin{table}
		\centering
		\begin{tabular}{llcccccc}
			\toprule
			& & \multicolumn{3}{c}{MAC with $\hat{\boldsymbol\psi}_1$} & \multicolumn{3}{c}{error factor of $\sigma_1/\sigma_2$}\\
			\cmidrule(lr){3-5}\cmidrule(lr){6-8}
			$\Rey_\tau$ & $AR$ & Cess & capped & capped & Cess & capped & capped\\
			& & $c=1$ & $c=1$ & (\ref{eq:selfconsistent_c}) & $c=1$ & $c=1$ & (\ref{eq:selfconsistent_c})\\
			\midrule
				$180$ & $2$ & $0.928$ & $0.951$ & $0.950$ & $1.07$ & $1.03$ & $1.06$\\
				 & $4$ & $0.831$ & $0.928$ & $0.916$ & $1.09$ & $1.11$ & $1.10$\\
				 & $8$ & $0.854$ & $0.948$ & $0.921$ & $1.47$ & $1.27$ & $1.11$\\
				\midrule
				$550$ & $2$ & $0.674$ & $0.845$ & $0.859$ & $1.11$ & $1.14$ & $1.15$\\
				 & $4$ & $0.725$ & $0.887$ & $0.894$ & $1.37$ & $1.32$ & $1.23$\\
				 & $8$ & $0.822$ & $0.931$ & $0.917$ & $1.78$ & $1.34$ & $1.18$\\
				\midrule
				$1000$ & $2$ & $0.689$ & $0.896$ & $0.899$ & $1.14$ & $1.14$ & $1.19$\\
				 & $4$ & $0.601$ & $0.879$ & $0.898$ & $1.25$ & $1.10$ & $1.11$\\
				 & $8$ & $0.812$ & $0.919$ & $0.919$ & $2.00$ & $1.50$ & $1.32$\\
				\midrule
				all & & $0.741$ & $0.900$ & $0.902$ & $1.31$ & $1.21$ & $1.17$\\
			\bottomrule
		\end{tabular}
		\caption{Autonomous prediction of the leading SPOD mode by the eddy-viscosity resolvent. Mean MAC of the leading left singular vector with $\hat{\boldsymbol\psi}_1$, and error factor of the gain separation against the SPOD value, the exponential of the mean of $|\ln[(\sigma_1/\sigma_2)/(\lambda_1/\lambda_2)^{1/2}]|$ over the waves of each combination and, in the last row, over all waves. Capped: profile (\ref{eq:capped_cess}) with $\theta=0.2$, at $c=1$ and at $c$ from (\ref{eq:selfconsistent_c}).}
		\label{tab:autonomous}
	\end{table}

	The capped profile gives a higher MAC than Cess in every wave at $c=1$ and in $80$ of the $83$ waves with $c$ from (\ref{eq:selfconsistent_c}), raising the overall mean from $0.74$ to $0.90$ with either choice of $c$ (table~\ref{tab:autonomous}). With the capped profile the mean MAC stays close to $0.9$ at every Reynolds number, whereas with Cess it falls from $0.87$ at $\Rey_\tau\approx180$ to $0.69$ at $1000$. Choosing the best $c$ for each wave would raise the overall mean only to $0.92$, so little is lost by fixing $c$ with (\ref{eq:selfconsistent_c}). The gain separation is more sensitive to $c$. Cess over-predicts $\sigma_1/\sigma_2$ increasingly with aspect ratio, by a factor of two at $AR=8$ and $\Rey_\tau\approx1000$. The capped profile at $c=1$ reduces this factor to $1.5$, and $c$ from (\ref{eq:selfconsistent_c}) reduces it further to $1.32$. The error factor over all waves falls accordingly. At $AR=2$, however, the error factor with $c$ from (\ref{eq:selfconsistent_c}) is slightly larger than at $c=1$.

	\begin{figure}
		\centering
		\includegraphics[width=\textwidth]{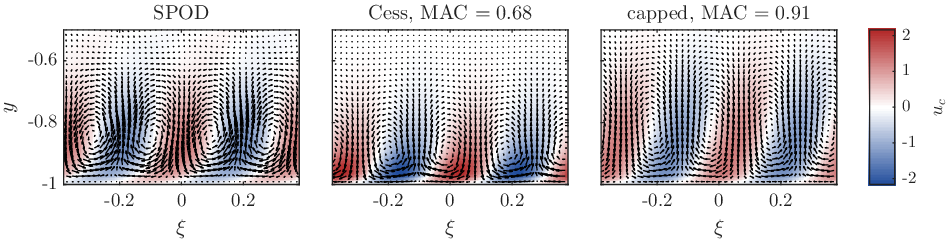}
		\caption{Plane normal to the wave crests, as in figure~\ref{fig:mode_reconstruction}, at $\Rey_\tau\approx550$, $AR=4$, $\lambda_x^+\approx850$, $\lambda_z^+\approx210$, for $-1\le y\le-0.5$. (a) Leading SPOD mode; (b) leading response mode of the eddy-viscosity resolvent with the Cess profile at $c=1$; (c) the model, with the capped profile (\ref{eq:capped_cess}) and $c$ from (\ref{eq:selfconsistent_c}). Colour is the component along $\boldsymbol e_c$, and the arrows are the in-plane vectors formed by the components along $\boldsymbol e_k$ and $\boldsymbol e_y$. Each mode has unit energy and is rotated into phase with the SPOD mode, and the colour and arrow scales are common to the three panels.}
		\label{fig:mode_reconstruction_model}
	\end{figure}

	Figure~\ref{fig:mode_reconstruction_model} shows the prediction for a sample wave at $\Rey_\tau\approx550$ and $AR=4$. As discussed before, the Cess profile overdamps the outer part of the streak (\S\ref{sec:model_profiles}) and confines the mode to the wall. The capped profile removes this overdamping, and its streak occupies the same height as in the SPOD mode. However, neither profile reproduces the rolls. In the SPOD mode the roll centres lie between the low- and high-speed streaks, which is the expected arrangement of lift-up. With the Cess profile the rolls are confined closer to the wall and, more importantly, their centres move towards the streak cores. The capped profile restores the position of the rolls relative to the streaks but extends them too far towards the channel centre. Both profiles also underpredict the energy of the rolls relative to that of the streak. The rolls hold $13$ per cent of the energy of the mode predicted with the Cess profile, against $27$ per cent in the SPOD mode. As for the streak, the capped profile reduces the roll overdamping and raises this share to $19$ per cent, which is still below the SPOD value. A possible reason is that $\mathrm{d}\mathsfbi V$ in (\ref{eq:Rev}) acts on the whole response, whereas the capped profile was built to approximate the null forcing alone. The next subsection examines this by restricting the eddy viscosity to the null sector.

	\subsection{The closed-loop resolvent for a given active forcing}
	\label{sec:model_closed_loop}

	When the active forcing is supplied, as it is when the measured RESPOD forcing is used, the eddy viscosity is needed only to represent the null forcing. In (\ref{eq:Rev}), however, it damps the whole response, including the part generated by the supplied active forcing. The sector decomposition suggests a different placement. Splitting the solenoidal forcing coefficients of (\ref{eq:fs_coeff}) as $\hat{\boldsymbol a}=\Pa\hat{\boldsymbol a}+\Pz\hat{\boldsymbol a}$ and applying $\Rs$ separates the response into the two contributions of (\ref{eq:sector_responses}),
	\begin{equation}
		\hat{\boldsymbol q}=\Rs\Pa\hat{\boldsymbol a}+\Rs\Pz\hat{\boldsymbol a}.
		\label{eq:cl_split}
	\end{equation}
	We retain the first term as an input. The second, the null contribution, was shown in \S\ref{sec:source_sink_group} to oppose the response and remove energy from it, and it is the only term that the eddy viscosity is used to represent here. In this subsection we absorb the coefficient $c$ of (\ref{eq:selfconsistent_c}) into $\mathrm{d}\mathsfbi V$, which is built from $c\,\delta\nu$ with the capped profile (\ref{eq:capped_cess}), so that $\Rev=(\Imat-\Rs\Bs^{\dagger}\,\mathrm{d}\mathsfbi V)^{-1}\Rs$ by (\ref{eq:Rev}). Modelling the null forcing as a diffusive response to the velocity field, $\hat{\boldsymbol f}_{0}\approx\mathrm{d}\mathsfbi V\hat{\boldsymbol q}$, its coefficients are $\Pz\hat{\boldsymbol a}\approx\Pz\Bs^{\dagger}\,\mathrm{d}\mathsfbi V\hat{\boldsymbol q}$, so that
	\begin{equation}
		\Rs\Pz\hat{\boldsymbol a}\approx\mathsfbi K\hat{\boldsymbol q},
		\qquad
		\mathsfbi K=\Rs\Pz\Bs^{\dagger}\,\mathrm{d}\mathsfbi V.
		\label{eq:cl_feedback}
	\end{equation}
	Substituting into (\ref{eq:cl_split}) gives $\hat{\boldsymbol q}=\Rs\Pa\hat{\boldsymbol a}+\mathsfbi K\hat{\boldsymbol q}$, and collecting the response on the left,
	\begin{equation}
		\hat{\boldsymbol q}\approx \Rcl\,\hat{\boldsymbol a},
		\qquad
		\Rcl=\big(\Imat-\mathsfbi K\big)^{-1}\Rs\Pa.
		\label{eq:closed_loop}
	\end{equation}
	The subscript denotes a closed loop. In this model the pressure-null sector no longer enters as an independent input but is returned to the operator as a feedback path, in which the response generates a null forcing through $\mathrm{d}\mathsfbi V$ and that forcing acts back on the response through $\Rs\Pz$. Because the response and the null forcing it generates are solved for together, $\Rcl$ maps the pressure-active forcing alone to the full response.

	Comparing (\ref{eq:Rev}) and (\ref{eq:closed_loop}), the two operators carry the same feedback force $\mathrm{d}\mathsfbi V\hat{\boldsymbol q}$. In $\Rev$ the whole force acts on the response, through $\Rs\Bs^{\dagger}$, whereas in $\Rcl$ only its null part does, through $\Rs\Pz\Bs^{\dagger}$. The null sector consists of forcings along the crests, so this part is the streak of $\mathrm{d}\mathsfbi V\hat{\boldsymbol q}$, which by (\ref{eq:dV}) is the diffusion of the streak $\hat u_c$ alone, and its response is a pure streak (\S\ref{sec:interference}). The feedback in $\Rcl$ therefore enters only the Squire equation. The roll is the response of the molecular Orr--Sommerfeld equation to the active forcing, as in $\Rs$, and the eddy viscosity damps only the streak that this roll drives by lift-up. In $\Rev$ the diffusion enters both equations, as in the eddy-viscosity operator of \citet{PalyaEtAl2026}. It damps the roll as well, which weakens the lift-up that drives the streak and yields the roll overdamping seen in figure~\ref{fig:mode_reconstruction_model}. Without the added diffusion both operators reduce to $\Rs$, so when they are driven by the same active forcing the placement of the dissipation is the only difference between them.

	The construction rests on the modelling assumption made in (\ref{eq:cl_feedback}), that the null response is approximated by the feedback,
	\begin{equation}
		\hat{\boldsymbol q}_{0}=\Rs\Pz\hat{\boldsymbol a}\;\approx\;\mathsfbi K\hat{\boldsymbol q}
		=\Rs\Pz\Bs^{\dagger}\,\mathrm{d}\mathsfbi V\hat{\boldsymbol q}.
		\label{eq:cl_approximation}
	\end{equation}
	We assess it through the response it produces. Both operators are driven by the pressure-active part $\Pa\hat{\boldsymbol a}$ of the leading RESPOD mode, and we compare their responses with the leading SPOD mode by the MAC and by the amplitude ratio $\|\hat{\boldsymbol q}\|/\|\hat{\boldsymbol\psi}_1\|$, which is unity when the model reproduces the measured magnitude.

	\begin{table}
		\centering
		\begin{tabular}{llcccccc}
			\toprule
			& & \multicolumn{2}{c}{MAC, measured forcing} & \multicolumn{2}{c}{MAC, white forcing} & \multicolumn{2}{c}{$\|\hat{\boldsymbol q}\|/\|\hat{\boldsymbol\psi}_1\|$}\\
			\cmidrule(lr){3-4}\cmidrule(lr){5-6}\cmidrule(lr){7-8}
			$\Rey_\tau$ & $AR$ & $\Rev$ & $\Rcl$ & $\Rev$ & $\Rcl$ & $\Rev$ & $\Rcl$\\
			\midrule
				$180$ & $2$ & $0.85$ & $0.91$ & $0.89$ & $0.85$ & $0.23$ & $0.70$\\
				 & $4$ & $0.90$ & $0.95$ & $0.90$ & $0.92$ & $0.37$ & $0.85$\\
				 & $8$ & $0.93$ & $0.97$ & $0.91$ & $0.95$ & $0.30$ & $0.82$\\
				\midrule
				$550$ & $2$ & $0.69$ & $0.84$ & $0.78$ & $0.62$ & $0.31$ & $0.72$\\
				 & $4$ & $0.86$ & $0.93$ & $0.88$ & $0.83$ & $0.43$ & $0.89$\\
				 & $8$ & $0.92$ & $0.97$ & $0.92$ & $0.92$ & $0.34$ & $0.95$\\
				\midrule
				$1000$ & $2$ & $0.64$ & $0.83$ & $0.81$ & $0.60$ & $0.31$ & $0.69$\\
				 & $4$ & $0.75$ & $0.89$ & $0.86$ & $0.67$ & $0.34$ & $0.81$\\
				 & $8$ & $0.92$ & $0.98$ & $0.92$ & $0.91$ & $0.37$ & $1.04$\\
				\midrule
				all & & $0.80$ & $0.91$ & $0.86$ & $0.77$ & $0.34$ & $0.82$\\
			\bottomrule
		\end{tabular}
		\caption{Eddy-viscosity resolvent $\Rev$ and closed loop $\Rcl$, both with the capped profile (\ref{eq:capped_cess}) and $c$ from (\ref{eq:selfconsistent_c}), against the leading SPOD mode. Measured forcing: mean MAC and mean amplitude ratio of the responses $\Rev\Pa\hat{\boldsymbol a}_1$ and $\Rcl\hat{\boldsymbol a}_1$. White forcing: mean MAC of the leading response mode when the forcing is white in the active sector. Means are over the waves of each combination and, in the last row, over all waves.}
		\label{tab:forced}
	\end{table}

	With the measured forcing, the closed loop gives a higher MAC than the eddy-viscosity resolvent in $82$ of the $83$ waves and raises the overall mean from $0.80$ to $0.91$ (table~\ref{tab:forced}). The improvement is largest where the eddy-viscosity resolvent has its lowest MAC, at $AR=2$ and the two higher Reynolds numbers. The difference in amplitude is larger. The eddy-viscosity resolvent recovers about a third of the measured amplitude, because it also damps the response to the active forcing, whereas the closed loop recovers $0.82$ of it on average, with ratios between $0.55$ and $1.12$.

	The white-forcing columns give the MAC of each operator when the forcing carries no information, as in the autonomous prediction of \S\ref{sec:model_autonomous}. The closed loop degrades with white forcing, its mean MAC reducing from $0.91$ to $0.77$, so the improved agreement with the SPOD mode comes from the use of the active forcing and not from the operator alone. With white forcing it is the weaker of the two operators, and the two are comparable only at $AR=8$. The eddy-viscosity resolvent behaves in the opposite way. With white forcing its mean MAC is slightly below the value of table~\ref{tab:autonomous}, because the restriction to the active sector removes the crest component of its optimal forcing, and it decreases further when the measured forcing is supplied. This is expected from the placement of the dissipation, since the eddy-viscosity resolvent also damps the roll that the measured active forcing drives, as discussed after (\ref{eq:closed_loop}). The low amplitude ratio shows this damping directly, and the weakened rolls of the autonomous mode in figure~\ref{fig:mode_reconstruction_model} show its effect on the shape.

	\begin{figure}
		\centering
		\includegraphics[width=.85\textwidth]{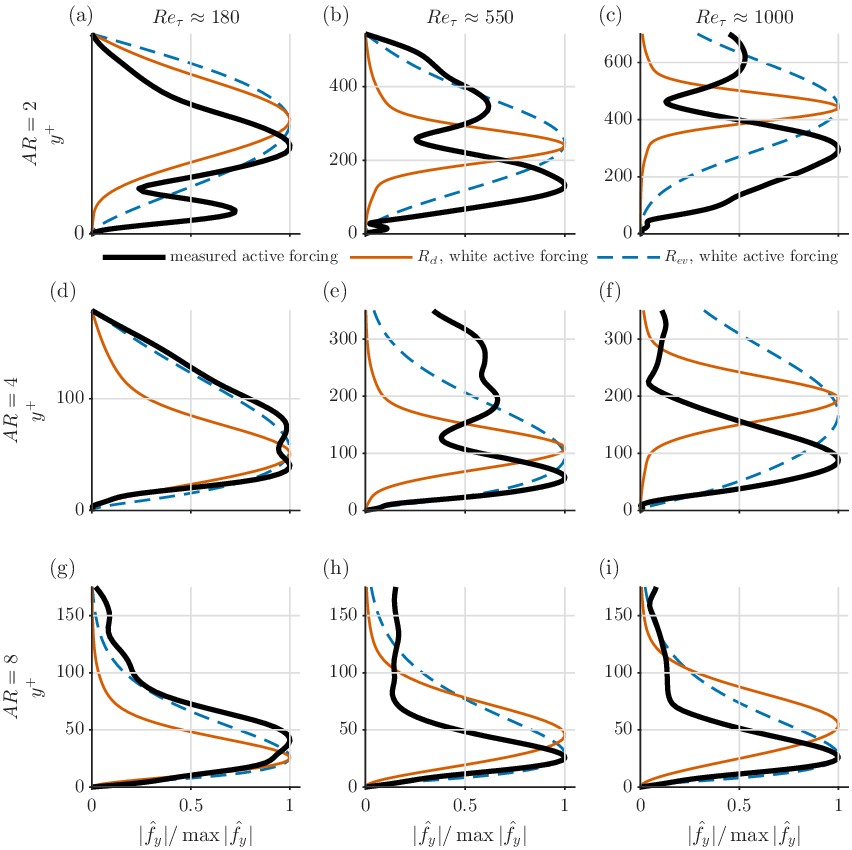}
		\caption{Wall-normal profile of $|\hat f_y|$ for the measured pressure-active forcing $\Pa\hat{\boldsymbol a}$ of the leading RESPOD mode, and for the leading forcing modes of $\Rcl$ and $\Rev$ with forcing white in the active sector, as in table~\ref{tab:forced}. One wave at $\lambda_x^+\approx1100$ in each combination: (a--c) $AR=2$, (d--f) $4$ and (g--i) $8$, with $\lambda_z^+\approx550$, $280$ and $140$, and $\Rey_\tau\approx180$, $550$ and $1000$ from left to right. Each profile is normalised by its maximum. The profiles extend to $y^+=700$, $350$ and $175$ for $AR=2$, $4$ and $8$, or to the channel centre where that is lower.}
		\label{fig:active_layers}
	\end{figure}

	Figure~\ref{fig:active_layers} suggests why the closed loop falls behind with white forcing. In the active sector the crest component of the forcing vanishes, and continuity gives $\hat f_k=(\ii/k)\,D\hat f_y$, so the wall-normal component $\hat f_y$ determines the active forcing completely. Each peak of $|\hat f_y|$ indicates the height of the centres of one layer of rolls. The white-forcing modes of both operators contain a single layer, here and in every wave, whereas the measured forcing often contains two or more, as in panels (a--c) and (e). Such waves are common at $AR=2$ and $4$, rarer at $AR=8$, and occur at every Reynolds number. The single layer of the models also lies at a different height. Where the measured forcing has several layers, it falls between them, and where it has one, it lies above or below it. White forcing cannot represent this structure, because each operator then selects its own optimal forcing. The share of multi-layer waves falls with aspect ratio, as the white-forcing deficit of the closed loop does. Within one aspect ratio, however, the layer count does not explain the deficit consistently, so the multi-layered rolls are a possible contributor to the deficit but not the only reason.

	With the measured forcing, the damping of the roll in $\Rev$ thus lowers both the MAC and the amplitude. With white forcing, in contrast, the eddy-viscosity resolvent gives the higher MAC, which suggests that the damping of the roll compensates in part for the colour of the active forcing that white forcing lacks. Such a compensation is consistent with the equivalence between coloured forcing and a modified operator described by \citet{ZareEtAl2017}. It is only partial, since the white-forcing mode of the eddy-viscosity resolvent also has a single layer of roll. Under this interpretation, the eddy viscosity of the autonomous model of \S\ref{sec:model_autonomous} represents both the null forcing on which it was calibrated and part of the colour of the active forcing.

	\section{Conclusions}
	\label{sec:conclusions}

		This study asked which part of the nonlinear forcing in a resolvent model of turbulent channel flow has to be supplied as an input and which part can be modelled through the response. To answer it, we divided the solenoidal forcing by the pressure it produces. The pressure-null sector produces no pressure-gradient response, and the pressure-active sector is its orthogonal complement. In a parallel flow and for $\alpha\neq0$, these are the Squire and Orr--Sommerfeld forcing subspaces. The pressure-active forcing therefore drives rolls, which generate streaks by lift-up, and the pressure-null forcing drives streaks directly. We applied the partition to the forcing associated with the leading SPOD mode of each wavenumber pair, extracted by RESPOD from DNS at $\Rey_\tau\approx180$, $550$ and $1000$.

		The two sectors have opposite roles in the energetic structures. Most of the forcing energy lies in the pressure-null sector, where the resolvent gain is low, whereas the standard resolvent selects forcing in the pressure-active sector, where the gain is several times higher. The pressure-active forcing supplies energy to the response and the pressure-null forcing removes it, and the streaks that the two sectors generate largely cancel. The streak of the leading structure is therefore the remainder of two opposing contributions. One is strongly amplified and comes from a small part of the forcing energy, and the other is weakly amplified and comes from most of it. Selection by amplification alone retains only the first. The opposition also varies with the instantaneous amplitude of the streak. Sorting the time histories by streak energy shows that the pressure-null forcing has nearly the same magnitude regardless of the streak being weak or strong. When the streak is strong, however, a larger part of this forcing is directed against the streak velocity, and it removes more energy per unit streak energy. This opposition is consistent with the stage of the self-sustaining process in which the nonlinear term opposes the streak, and it is observed here in unforced turbulence at every wavenumber pair examined.

		These roles indicate that the two sectors require different representations in a model. The pressure-null forcing is related to the response as a diffusion acting on the streak would be. It opposes the streak in phase and removes energy at a rate that, relative to molecular dissipation, grows with the wavelength. This motivates modelling it as an eddy-viscosity feedback. The Cess profile commonly used for this purpose was designed to reproduce the mean Reynolds shear stress. Here we instead chose the eddy viscosity so that the forcing it implies matches the measured pressure-null forcing. The result is a Cess profile capped above a height proportional to the spanwise wavelength, with an amplitude set by one relation for all wavenumber pairs. The resolvent built with this eddy viscosity predicts the leading SPOD mode more closely than the Cess profile does, and the improvement is largest at the highest Reynolds number. When the pressure-active forcing, which holds about $10$ per cent of the forcing energy, is taken from the DNS and the remaining forcing is modelled with this eddy viscosity, we reconstruct the shape of the leading SPOD modes with a mean modal assurance criterion of $0.91$ and recover $82$ per cent of their amplitude, on average over all wavenumber pairs. This framework is not yet predictive, however, since it requires the pressure-active forcing as an input. The measured pressure-active forcing often contains several layers of rolls, and modelling this structure requires further investigation. Only its wall-normal component has to be modelled, since the remaining components follow from it through the solenoidal constraint. Even with a model of the pressure-active forcing, the framework would remain linear, with an eddy viscosity that is fixed in time. It therefore cannot represent the stronger alignment of the pressure-null forcing against the streak at instants of high streak energy. Its predictions are thus limited to the second-order statistics of the response and do not extend to individual realisations.

		Finally, the pressure-based partition is not restricted to the channel. Its definition uses only the pressure response of the resolvent and a basis of solenoidal forcing, so the two sectors can be computed for the global resolvent operators of non-parallel flows, such as jets and boundary layers, where no Orr--Sommerfeld/Squire decomposition exists. A similar forcing partition can then be examined in these flows. Extending the definition of the sectors to compressible flows, where the forcing is no longer solenoidal, is a further direction for future work.

		\begin{bmhead}[Acknowledgements.]
	The author thanks Andr\'e V. G. Cavalieri for providing the DNS databases and for insightful comments on this work.
	\end{bmhead}

	\begin{bmhead}[Declaration of AI use.]
	During the preparation of this work the author used an AI assistant, Claude (Anthropic; model Claude Fable 5.2 and earlier versions, accessed through the Claude Science application), to draft and revise parts of the text, to reorganise the manuscript, and to assist in writing and testing the scripts used for the data analysis and the figures. The tool was also used to check the reported numbers against the data and the citations against their sources. The author reviewed and verified all text, code and results, and takes full responsibility for the content of the paper.
	\end{bmhead}

	\begin{bmhead}[Funding.]
	This research received no specific grant from any funding agency.
	\end{bmhead}

	\begin{bmhead}[Declaration of interests.]
	The author reports no conflict of interest.
	\end{bmhead}

	\begin{bmhead}[Data availability statement.]
	The DNS data that support the findings of this study are available from the author upon request.
	\end{bmhead}

	\begin{bmhead}[Author ORCIDs.]
	U. Karban, https://orcid.org/0000-0003-0803-0581.
	\end{bmhead}

	\begin{appen}
	\section{Construction of the forcing sectors}
	\label{app:sectors}
	This appendix details the discrete construction of the two forcing sectors defined in \S\ref{sec:theory}.
	\subsection{Solenoidal basis}

 The fields are discretised in the wall-normal direction on the Chebyshev points on which the DNS data are stored. A forcing field is represented by the values of its three components at these points, and $\Wmat$ holds the Clenshaw--Curtis quadrature weights. Because the forcing, the velocity response and the pressure-gradient response are all three-component fields on the same grid, the same matrix measures each of them.

	The basis $\Bs$ is constructed in two steps. The forcing is first restricted to fields that vanish at the two walls in all three components.
	The fields are also restricted to one symmetry class about the channel centre, with the streamwise and spanwise components even and the wall-normal component odd. The columns of $\Bs$ then span the kernel of the discrete divergence operator $[\ii\alpha\Imat,\ \Dmat,\ \ii\beta\Imat]$ on these fields, and are orthonormalised with respect to $\Wmat$,
	\begin{equation}
		\Bs^{*}\Wmat\Bs=\Imat.
		\label{eq:Bs_orthonormal}
	\end{equation}
	With this normalisation, the squared Euclidean norm of the coordinates $\hat{\boldsymbol a}$ of (\ref{eq:Bs_dagger}) is the forcing energy,
	\begin{equation}
		\hat{\boldsymbol a}^{*}\hat{\boldsymbol a}
		=
		\hat{\boldsymbol f}_{\mathrm{s}}^{*}\Wmat\hat{\boldsymbol f}_{\mathrm{s}}.
		\label{eq:energy_coordinates}
	\end{equation}
	An orthogonal projection of the coordinates therefore divides the forcing energy into parts that add up.

	\subsection{Forcing subspaces and the pressure they generate}

	With the Orr--Sommerfeld forcing map
	\begin{equation}
		\mathsfbi L_{\mathrm{OS}}
		=
		\begin{bmatrix}
			-\ii\alpha\Dmat & -k^{2}\Imat & -\ii\beta\Dmat
		\end{bmatrix},
		\label{eq:Los}
	\end{equation}
	the Squire forcing subspace is the set of forcing coordinates that produce no Orr--Sommerfeld forcing, $\mathcal F_{\mathrm{Sq}}=\ker(\mathsfbi L_{\mathrm{OS}}\Bs)$, and the Orr--Sommerfeld forcing subspace is its orthogonal complement, $\mathcal F_{\mathrm{OS}}=\mathcal F_{\mathrm{Sq}}^{\perp}$. Because $\Bs$ is $\Wmat$-orthonormal, orthogonality of the coordinates is orthogonality in forcing energy.

	For incompressible perturbations about a parallel mean flow, with divergence-free forcing, the pressure-Poisson equation is
	\begin{equation}
		\Delta\hat p=-2\ii\alpha U'\hat v,
		\qquad
		U'=\mathrm{d}U/\mathrm{d}y.
		\label{eq:poisson}
	\end{equation}
	The pressure in (\ref{eq:poisson}) is the rapid pressure of the rapid--slow decomposition \citep{Kim1989,LuharEtAl2014}. The slow pressure is generated by the irrotational part of the forcing, which produces no velocity, and is therefore absent here. A forcing with $\hat g_{\mathrm{OS}}=0$ has $\hat f_y=0$ by (\ref{eq:gos_fy}) and produces no wall-normal velocity. With $\hat v=0$ and $\hat f_y=0$, the wall-normal momentum equation reduces to $D\hat p=0$. Equation~(\ref{eq:poisson}) then gives $k^{2}\hat p=0$, so the pressure vanishes and there is no pressure-gradient response. Conversely, for $\alpha\neq0$, a zero pressure gradient implies $\hat p=0$, because its streamwise component is $\ii\alpha\hat p$. Equation~(\ref{eq:poisson}) then gives $U'\hat v=0$, hence $\hat v=0$ because $U'$ vanishes only at the channel centre, and the Orr--Sommerfeld equation in turn yields $\hat g_{\mathrm{OS}}=0$.

	\subsection{Singular value decomposition of the pressure operator}

	The full singular value decomposition of $\Js$ in (\ref{eq:Js}) is partitioned as
	\begin{equation}
		\Js
		=
		\Uj
		\begin{bmatrix}
			\Sa & \mathsfbi{0}
		\end{bmatrix}
		\begin{bmatrix}
			\Na & \Nz
		\end{bmatrix}^{*},
		\label{eq:Js_svd}
	\end{equation}
	where $\Sa$ contains the non-zero singular values. The columns of $\Nz$ span the kernel $\mathcal N_0$, so that $\Js\Nz=\mathsfbi{0}$, and the columns of $\Na$ span its orthogonal complement $\mathcal N_{\mathrm a}$. Measuring the pressure-gradient response in the norm (\ref{eq:L2_norm}) would change the non-zero singular values of $\Js$ but not its kernel, so the sectors do not depend on this choice.

	\section{Spectral estimation and modal decompositions}
	\label{app:modes}
This appendix gives the window used in the spectral estimation of \S\ref{sec:database} and the definitions of the SPOD and RESPOD modes introduced in \S\ref{sec:spod_respod}.
	\subsection{Window}

	Each block is multiplied by the window
	\begin{equation}
		w(\tau)=\mathrm{e}^{8}\exp\left\{-\frac{2}{\tau(1-\tau)}\right\},
		\label{eq:window}
	\end{equation}
	where $\tau\in[0,1]$ is the normalised time within the block, and is then Fourier transformed. This is the infinitely smooth window of \citet{MartiniEtAl2019arXiv}, with their parameter $n=2$. Windowing adds a forcing-like term, the time derivative of the window multiplied by the response, to the linearised equations \citep{MartiniEtAl2019arXiv}. We add its Fourier transform to each forcing block, so that the resolvent relates the windowed forcing and response exactly.

	\subsection{SPOD and RESPOD modes}

	We collect the realisations $\hat{\boldsymbol q}^{(k)}$ and $\hat{\boldsymbol a}^{(k)}$ of \S\ref{sec:spod_respod} as the columns of
	\begin{equation}
		\hat{\mathsfbi Q}=(N-1)^{-1/2}\big[\hat{\boldsymbol q}^{(1)},\dots,\hat{\boldsymbol q}^{(N)}\big],
		\qquad
		\hat{\mathsfbi A}=(N-1)^{-1/2}\big[\hat{\boldsymbol a}^{(1)},\dots,\hat{\boldsymbol a}^{(N)}\big],
		\label{eq:realisation_matrices}
	\end{equation}
	so that $\hat{\mathsfbi Q}\hat{\mathsfbi Q}^{*}$ and $\hat{\mathsfbi A}\hat{\mathsfbi A}^{*}$ are the response and forcing cross-spectral densities. SPOD \citep{TowneSchmidtColonius2018} decomposes the response into modes that are orthogonal in the $L^2$ norm and ranked optimally by the mean fluctuation energy they capture. The modes are the eigenvectors of $\hat{\mathsfbi Q}\hat{\mathsfbi Q}^{*}\Wmat$ and follow from the singular value decomposition of the realisations, taken in the norm (\ref{eq:L2_norm}),
	\begin{equation}
		\hat{\mathsfbi Q}=\hat{\boldsymbol\Psi}\boldsymbol\Lambda^{1/2}\boldsymbol\Phi^{*}.
		\label{eq:spod}
	\end{equation}
	The columns $\hat{\boldsymbol\psi}_j$ of $\hat{\boldsymbol\Psi}$, which satisfy $\hat{\boldsymbol\Psi}^{*}\Wmat\hat{\boldsymbol\Psi}=\Imat$, are the SPOD modes, and the diagonal matrix $\boldsymbol\Lambda$ holds their energies $\lambda_1\geq\lambda_2\geq\cdots$. Each column $\boldsymbol\phi_j$ of $\boldsymbol\Phi$ holds the weights with which the realisations combine into the mode, $\hat{\boldsymbol\psi}_j=\lambda_j^{-1/2}\hat{\mathsfbi Q}\boldsymbol\phi_j$.

	RESPOD \citep{KarbanEtAl2022} is the extended POD \citep{Boree2003} of the forcing with respect to these modes. It applies the same weights to the forcing realisations,
	\begin{equation}
		\hat{\boldsymbol a}_j=\lambda_j^{-1/2}\hat{\mathsfbi A}\boldsymbol\phi_j.
		\label{eq:respod}
	\end{equation}
	Every realisation satisfies $\hat{\boldsymbol q}^{(k)}=\Rs\hat{\boldsymbol a}^{(k)}$, so $\hat{\mathsfbi Q}=\Rs\hat{\mathsfbi A}$. Applying $\Rs$ to (\ref{eq:respod}) then gives (\ref{eq:respod_closure}).

	Since $\hat{\mathsfbi Q}=\Rs\hat{\mathsfbi A}$, (\ref{eq:spod}) is also the singular value decomposition of $\Rs\hat{\mathsfbi A}$, the resolvent coloured by the forcing statistics, whose input singular vectors are the weights $\boldsymbol\phi_j$. By (\ref{eq:respod}), the RESPOD forcing modes are these input singular vectors mapped into forcing coordinates by $\hat{\mathsfbi A}$ and scaled by $\lambda_j^{-1/2}$. For white forcing with $\hat{\mathsfbi A}\hat{\mathsfbi A}^{*}=\Imat$, the vectors $\hat{\mathsfbi A}\boldsymbol\phi_j$ are the right singular vectors of $\Rs$, so the RESPOD modes reduce to the forcing modes of the standard resolvent divided by their gains.
\end{appen}

	\bibliographystyle{jfm}
	\bibliography{jfm}

\end{document}